\documentclass[12pt]{article}
\usepackage{amssymb,amsmath}
\usepackage{latexsym}
\usepackage{graphicx}
\usepackage{subfigure}
\usepackage{booktabs}
\usepackage{bbm}
\usepackage{enumitem}
\usepackage{color}
\usepackage{datetime}
\usepackage{cite}
\usepackage[margin=20pt,small]{caption}
\usepackage{psfrag}
\usepackage[
      colorlinks=false,
      linkcolor=darkblue,  
      urlcolor=blue,    
      filecolor=blue,     
      citecolor=red,
linktocpage=true,
      pdfstartview=FitV,
      bookmarksopen=true    
      ]{hyperref}

\let\oldbibliography\thebibliography
\renewcommand{\thebibliography}[1]{\oldbibliography{#1}
\setlength{\itemsep}{0pt}} 

\numberwithin{equation}{section}  

\makeatletter
\g@addto@macro\bfseries{\boldmath}
\makeatother

\definecolor{cardinal}{rgb}{0.6,0,0}
\definecolor{darkgreen}{rgb}{0,0.5,0}
\definecolor{golden}{rgb}{0.92, 0.7, 0}
\definecolor{midnight}{rgb}{0, 0, 0.5}
\definecolor{darkblue}{rgb}{0.2, 0, 0.8}
\newcommand{\Red}{\color{red}}

\def\NPWc#1{{\Red{\bf NPW:} {\tt #1}}}

\def\IC{\mathbb{C}}
\def\IP{\mathbb{P}}
\def\RR{\mathbb{R}}
\def\CC{\mathbb{C}}
\def\PP{\mathbb{P}}

\def\Neql#1{{\cal N}\!=\!{#1}}

\def\eql{{~=~}}
\def\coeff#1#2{\relax{\textstyle {#1 \over #2}}\displaystyle}

\def\IR{\mathbb{R}}
\def\ZZ{\mathbb{Z}}

\def\cB{{\cal B}}

\def\cI{{\cal I}}

\def\cM{{\cal M}}

\def\nBPS#1{$\frac{1}{#1}$-BPS}

\def\cO{{\cal O}}

\def\bfs#1{{\mathbf{#1}}}
\def\cals#1{{\mathcal{#1}}}

\def\eo{\overset{_{\phantom{.}\circ}}{e}{}}
\def\go{\overset{_{\phantom{.}\circ}}{g}{}}
\def\Do{\overset{_{\phantom{.}\circ}}{D}{}}

\def\Gao{\overset{_{\circ}}{\Gamma}{}}

\def\zetao{\overset{_{\circ}}{\zeta}{}}
\def\sto{\overset{_{\circ}}{\ast}}
\def\oo{\overset{_{\circ}}{o}{}}

\def\DX{X}
\def\omg{\upsilon}
\def\varth{\omega}

\def\til{~\sim~}

\def\fB{\frak B}
\def\fA{\frak A}

\def\suthuu{SU(3)$\times$\rm U(1)$\times$U(1)}

\begin{document}

\begin{titlepage}

\bigskip
\bigskip
\bigskip
\centerline{\LARGE \bf  $\Neql2$ Supersymmetric Janus Solutions and Flows:}
\medskip
\centerline{\LARGE \bf  From Gauged Supergravity to M Theory}

\bigskip\bigskip\bigskip

\centerline{{\bf Krzysztof Pilch, Alexander Tyukov   and Nicholas P. Warner}}
\bigskip

\centerline{Department of Physics and Astronomy,}
\centerline{University of Southern California,} 
\centerline{Los Angeles, CA 90089, USA}
\bigskip
\bigskip
\centerline{{\rm pilch@usc.edu,~tyukov@usc.edu,~warner@usc.edu} }
\bigskip
\bigskip

\bigskip\bigskip

\begin{abstract}
\noindent 
We investigate a family of SU(3)$\times$U(1)$\times$U(1)-invariant holographic flows and Janus solutions obtained from gauged $\Neql 8$ supergravity in four dimensions.  We give complete details of how to use the uplift formulae to obtain the corresponding solutions in M theory.    While the flow solutions  appear to be singular from the four-dimensional perspective, we find that the eleven-dimensional solutions are much better behaved and give rise to interesting new classes of compactification geometries that are smooth, up to orbifolds, in the infra-red limit.  Our solutions involve new phases in which M2 branes polarize partially or even completely into M5 branes.    We derive the eleven-dimensional supersymmetries and show that the eleven-dimensional equations of motion and BPS equations are indeed satisfied as a consequence of their four-dimensional counterparts. Apart from elucidating a whole new class of eleven-dimensional Janus and flow solutions, our work provides extensive and highly non-trivial tests of the recently-derived uplift formulae.

\end{abstract}

\end{titlepage}


\tableofcontents


\section{Introduction}
\label{Sect:introduction}

Finding and understanding the infra-red structure of holographic RG flows of $\Neql4$ Yang-Mills theory and of ABJM theory \cite{Aharony:2008ug} remains an immensely rich but rather challenging subject that still has the capacity to surprise.   In this context, gauged $\Neql8$ supergravity in four and five-dimensions has  proven to be a very powerful tool that continues to be extremely useful and yield interesting new results.

The catalog of physically interesting holographic solutions and flows that have been explicitly constructed in gauged supergravity is a very long one, whose early examples included  the flows to highly non-trivial $\Neql1$ supersymmetric ``Leigh-Strassler''  fixed points   \cite{Khavaev:1998fb,Freedman:1999gp,Pilch:2000ej,Pilch:2000fu,Halmagyi:2005pn} and its ABJM analog  (\cite{Warner:1983vz,Corrado:2001nv,Benna:2008zy} and  \cite{Ahn:2000aq,Ahn:2000mf,Ahn:2001by,Ahn:2001kw,Ahn:2002eh,Ahn:2002qga}), through examples of  $\Neql2$ Seiberg-Witten  flows  \cite{Pilch:2000ue,Buchel:2000cn,Evans:2000ct}, to maximally supersymmetric flows \cite{Pope:2003jp,Bena:2004jw,Lin:2004nb}.  There are many more examples, perhaps the most recent of which is the supersymmetric flow that we studied in  \cite{Pilch:2015vha}, where the large-$N$ theory on a stack of M2 branes flows to a new, ``nearly conformal'' supersymmetric theory in $(3+1)$ dimensions.  One of our purposes here is to discuss these new families of flows and related Janus solutions in some detail.  Another purpose of this paper is to highlight and explain some of the new techniques that were used in \cite{Pilch:2015vha}.

As has been noted in many places (see, for example, \cite{Freedman:1999gk,Warner:2000qh,Gubser:2000nd,Gowdigere:2005wq,Pilch:2015vha}), finding a holographic flow solution in lower dimensions, perhaps  in some consistent truncation, does not often give direct insight into the underlying physics.  More specifically, a holographic flow in the lower-dimensional theory may be singular and fields may flow to infinite values.  It is only when these solutions are uplifted to M theory or IIB supergravity that one can give a proper interpretation of the singular behavior in terms of a distribution of branes and fluxes. In this way, singular low-dimensional solutions may actually encode very interesting physics in higher dimensions.

This, of course, raises the obvious question as to why one does not simply start in the higher-dimensional theory from the outset.  The answer is straightforward: the lower-dimensional theory encodes fields much more simply and computably; complicated fluxes and metric deformations on internal manifolds become scalars described in terms of a potential.  The supersymmetries may also involve these internal fluxes and geometry in non-trivial ways.  The practical algorithms for solving supersymmetry variations directly in higher-dimensions therefore typically require the imposition of a high level of symmetry and supersymmetry.  The power of using the low-dimensional theory and its potential structure is that one can handle solutions that have much lower levels of symmetry and supersymmetry.   As will become evident, finding the solutions that we construct here directly in M theory is a truly daunting task, even when one knows exactly where to look.  

The price of working in lower-dimensional gauged supergravity is that it describes a very restricted family of deformations. On the other hand they are some of the most interesting deformations since they are dual to marginal and relevant operators.  What is surprising is that even after fifteen years since the first holographic flows in gauged supergravity \cite{Girardello:1998pd,Distler:1998gb,Freedman:1999gp}, there are still interesting new physical flows to be found (like the one in \cite{Pilch:2015vha}) and new Janus solutions that can be constructed explicitly in gauged supergravity  (like those in \cite{Bobev:2013yra}).   

Since gauged supergravity continues to give us new very interesting, physical solutions, while their interpretation usually requires the ``uplift'' to M theory or IIB supergravity,  it becomes ever more important to understand and develop the precise relationships between the gauged supergravity in low dimensions and the higher-dimensional supergravities.  In particular, one wants to develop explicit uplift formulae that provide the exact M theory of IIB solution in terms of the gauged supergravity fields.  There is also a vast literature on this subject and there has also been some remarkable progress on this in the last two or three years.  For simplicity, we will only give a very brief review here and restrict our attention entirely to M theory and its relation to gauged $\Neql8$ supergravity in four dimensions.

Gauged $\Neql8$ supergravity in four dimensions was first constructed in \cite{de Wit:1982ig} and there was a great deal of subsequent work that argued how this must be related to the $S^7$-compactification of M theory. (For a review, see \cite{Duff:1986hr}.)  There was extensive discussion as to whether gauged supergravity was a consistent truncation or merely a low-energy effective field theory. The former is a much stronger statement in that it means that solving the equations of motion in gauged supergravity {\it guarantees} that one has also solved the equations of motion of the higher dimensional theory.  Over the years it has become evident that 
the gauged theory is indeed a consistent truncation and formulae have emerged showing precisely how gauged supergravity encodes solutions to M theory.  

One of the first general formula was given in \cite{deWit:1984nz} where it was shown  how to compute the exact deformed metric on the $S^7$ in terms of all the supergravity scalars. This knowledge alone was immensely useful in finding uplifted solutions explicitly, see for example \cite{Corrado:2001nv,Ahn:2001kw,Ahn:2002eh}.   Exact formulae for fluxes proved to be a much greater challenge.    Indeed, the formulae for the components of the 4-form field strength obtained as part of the original proof of consistent truncation of M~theory on $S^7$ in \cite{deWit:1986mz,deWit:1986iy} were  prohibitively difficult to use and also suffered from an ambiguity  that would lead to some components having a wrong symmetry \cite{Nicolai:2011cy}. It is only recently that a new set of considerably more workable uplift Ans\"atze for the internal 3-form potential have been proposed in  \cite{deWit:2013ija,Godazgar:2013nma} and then extended to the other  components of the flux \cite{Godazgar:2013dma,Godazgar:2013pfa,Godazgar:2014nqa,Godazgar:2015qia}. However, explicit tests of those new formulae \cite{Godazgar:2013nma,Godazgar:2013pfa,Godazgar:2014eza} were confined to uplifts of simplest solutions of four-dimensional $\Neql 8$ supergravity, namely the $AdS_4$ solutions for the stationary points of the potential. Hence, it is important to perform non-trivial tests for solutions with varying scalar fields, such as  holographic flows.

 In looking at the holographic flows described in this paper and in \cite{Pilch:2015vha}, we first tried to find the uplift based entirely on knowing the internal metric through the formula of \cite{deWit:1984nz}.  This turned out to be impossible and the flux uplift formula became an essential part of constructing the flow solution. Moreover, the Janus solutions are intrinsically even more complicated and we certainly could not have constructed them without knowing how to uplift the fluxes.  We find that the new formula to uplift the fluxes \cite{deWit:2013ija,Godazgar:2013nma,Godazgar:2013dma,Godazgar:2013pfa,Godazgar:2014nqa,Godazgar:2015qia} do indeed generate the exact solution.  The only down-side is that they involve some heavy computations to arrive at a relatively simple result.

While our focus in this paper will be mainly on the details of how to construct the uplifts, one should not lose sight of the interesting physics of the solutions that we construct.  As we described in \cite{Pilch:2015vha}, our flow solutions start from a UV fixed point of M2 branes that, under a relevant perturbation, go to solutions sourced largely, or even entirely by M5 branes in the IR.  Unlike many flows to the IR, these flows, when uplifted, have only mild orbifold singularities.  Moreover, there is a special class of flows that go to pure M5 branes and can be interpreted as describing an ``almost conformal'' fixed point in $(3+1)$ dimensions.  This was the focus of our earlier paper \cite{Pilch:2015vha}.  In this paper we will also look at Janus solutions that delve into the backgrounds described by the flows and so may be interpreted as describing interfaces between phases described by the holographic IR flows.

To date, much of the discussion of Janus solutions  has been done directly in IIB supergravity  \cite{Bak:2003jk,Clark:2004sb,D'Hoker:2006uu,D'Hoker:2007xy} (see, however, \cite{Clark:2005te,Suh:2011xc}) or   M theory   \cite{D'Hoker:2008wc,D'Hoker:2008qm,D'Hoker:2009gg,Estes:2012vm,Bachas:2013vza}. As we remarked above, such direct constructions usually require a high level of symmetry, or supersymmetry to make  the computations feasible.  In particular, until  \cite{Bobev:2013yra}, there were very few \nBPS{4} (and no \nBPS{8}) Janus solutions in  M theory known.  These new Janus solutions were obtained in gauged supergravity and so we want to take one of the most non-trivial families of such flows and uplift to M theory so as to reveal the underlying geometric structure.

In section~\ref{Sect:truncation} we describe sector of gauged $\Neql8$ supergravity upon which we will focus and describe the BPS flow and Janus solutions from the four-dimensional perspective.  In section~\ref{Sect:uplift} we give the details of how this sector of gauged supergravity uplifts to M theory, while in  section~\ref{Sect:EOMS} we show how this uplifted solution solves the equations of motion in M theory. The supersymmetry structure of the solutions is studied from the eleven-dimensional perspective in section~\ref{Sec:susy11}.  In section \ref{Sec:LargeLambda} we return to studying the IR limits of our holographic flows and how they are related to distributions of M5 and M2 branes. We discuss general features of our solutions and how one might obtain more general families of solutions in
section \ref{Sect:General}.  Section \ref{Sect:conclusions} contains our concluding remarks. Our conventions and the tabulation of some of the more complicated formulae are given in the appendices.

\section{The truncation and BPS equations in four dimensions}
\label{Sect:truncation}

\subsection{The truncation}
\label{subSec:trunc}

In this section we summarize some explicit results  for the   truncation  of four-dimensional, $\Neql8$ supergravity~\cite{de Wit:1982ig} to the \suthuu-invariant sector that we will  need  for the uplift to M theory in section~\ref{Sect:uplift}.  Our discussion here is based on   \cite{Bobev:2013yra} and \cite{Pilch:2015vha}. The Lagrangian for the truncation   can also be read-off from a more general $\rm SU(3)$-invariant truncation  in  \cite{Warner:1983vz} and \cite{Bobev:2009ms,Bobev:2010ib}.

The   \suthuu$\,\rm \subset SO(8)$ symmetry group of the truncation is defined  by its action on the supersymmetries, $\epsilon^i$, of the $\Neql8$ theory.    We choose $\rm SU(3)$ and the first $\rm U(1)$ to act on the indices $i=1,\ldots,6$, while the second $\rm U(1)$  on the indices $i=7, 8$. This corresponds to the branching 
\begin{equation}\label{8vbranch}
\bfs 8_v\quad\longrightarrow\quad (\bfs 3, 1,0)+(\overline{\bfs 3},-1,0)+(\bfs 1,0,1)+(\bfs 1,0,-1)\,.
\end{equation}
The resulting truncation is particularly simple since, as observed in \cite{Bobev:2013yra},  the  commutant of the symmetry group in $\rm E_{7(7)}$ consists   of a single $\rm SL(2,\RR)$. The invariant fields are: the graviton, $g_{\mu\nu}$, the gauge field, $A_\mu^\alpha$, for the two $\rm U(1)$'s, a scalar, $x$, and a pseudoscalar, $y$.  As we will describe below, this may be viewed as the bosonic sector of $\Neql 2$ supergravity coupled to a vector multiplet.

The  two non-compact generators of  $\rm SL(2,\RR)$ in the fundamental representation of  $\rm E_{7(7)}$ can be chosen as follows:
\begin{equation}\label{Tgens}
\bfs T_s\eql \left(\begin{matrix}
0 & \Phi^+_{IJKL}\\
\Phi^+_{IJKL} & 0
\end{matrix}\right)\,,\qquad \bfs T_c\eql \left(\begin{matrix}
0 & i\,\Phi^-_{IJKL}\\
-i \,\Phi^-_{IJKL} & 0
\end{matrix}\right)\,,
\end{equation}
where
\begin{equation}\label{Ytens}
\Phi^\pm _{IJKL}\eql 24\,(\delta^{1234}_{IJKL}+\delta^{1256}_{IJKL}\pm \delta^{1278}_{IJKL}+\delta^{3456}_{IJKL}\pm\delta^{3478}_{IJKL}\pm \delta^{5678}_{IJKL})\,,
\end{equation}
are  self-dual ($+$) and  antiself-dual ($-$) $\rm SO(8)$ tensors, respectively. 
Then the scalar `56-bein'  is
\begin{equation}\label{56bein}
\cals V ~\equiv~ e^{\bfs V}\eql \left(\begin{matrix}
u_{ij}{}^{IJ} & v_{ijIJ}\\ v^{ijIJ} & u^{ij}{}_{IJ}
\end{matrix}\right)\,,\qquad \bfs V \eql x\,\bfs T_s+y\,\bfs T_c\,,
\end{equation}
where the scalar, $x\equiv\lambda\cos\zeta$, and the pseudoscalar, $y\equiv\lambda\sin\zeta$, parametrize the coset
\begin{equation}\label{sl2coset}
\rm {SL(2,\RR)\over SO(2)}\,,
\end{equation}
with the canonical complex coordinate, $z$, given by 
\begin{equation}\label{}
z \eql\tanh\lambda\, e^{i\zeta}\,.
\end{equation}
Given the explicit generators \eqref{Tgens}, it is easy to check that the exponential \eqref{56bein} reduces to a polynomial,
\begin{equation}\label{expV}
\cals V\eql a_0+a_1\,\bfs V+a_2\,\bfs V^2+a_3\,\bfs V^3\,,
\end{equation}
where
\begin{equation}\label{}
a_0\eql {2-3|z|^2\over 2(1-|z|^2)^{3/2}}\,,\qquad a_1\eql {6-7|z|^2\over 6(1-|z|^2)^{3/2}}\,,\qquad a_2\eql 3\,a_3\eql {1\over  2(1-|z|^2)^{3/2}}\,,
\end{equation}
Note that the order of this polynomial coincides with the index of embedding of $\rm SL(2,\RR)$ in $\rm E_{7(7)}$.

Using the  56-bein \eqref{56bein}, it is now straightforward to   obtain the full bosonic action of the truncated theory  \cite{Bobev:2010ib}. In particular, we find that it is consistent to set the vector fields, $A_\mu^\alpha$, to zero. Then the resulting Lagrangian for the gravity coupled to the  scalar fields is:\footnote{See, \cite{Bobev:2009ms,Bobev:2010ib} and appendix~\ref{Appendix:B}.} 
\begin{equation}\label{theLag}
\begin{split}
e^{-1}{\cals L}& \eql {1\over 2}\,R -3\,{\partial_\mu z\partial^\mu\bar z\over (1-|z|^2)^2}-6g^2\,{1+|z|^2\over 1-|z|^2}\\[6 pt]
& \eql {1\over 2}\,R   -3\,\partial_\mu\lambda\partial^\mu\lambda -{3\over 4}\sinh^2(2\lambda)\,\partial_\mu\zeta\partial^\mu\zeta+6g^2\cosh(2\lambda)\,.
\end{split}
\end{equation}

The Lagrangian \eqref{theLag} has no explicit dependence on the phase, $\zeta$,   and hence there is a conserved Noether current 
\begin{equation}\label{consJ}
\cals J_\mu\eql e\,\sinh^2(2\lambda)\,\partial _\mu\zeta\,,
\end{equation}
with the corresponding $U(1)_\zeta$ symmetry being simply a rotation between the scalar and the pseudoscalar.

It was shown in \cite{Bobev:2010ib,Bobev:2013yra} that by keeping the $\rm SU(3)$-invariant fermions, the truncation yields a  $\Neql 2$  supergravity in four dimensions. Its   $R$-symmetry  is a combination of the two $U(1)$'s and, from the supersymmetry variations,
\begin{equation}\label{}
\delta\psi^{i}_\mu\eql 2 D_\mu\epsilon^i+\sqrt 2\,g\,A_a{}^{ij}\gamma_\mu\epsilon_j\,,\qquad i,j=7,8\,,
\end{equation}
the real superpotential, $W$, is given by an eigenvalue of the $A_1$-tensor,
$W=\sqrt 2\,|A_1{}^{77}|=\sqrt 2\,|A_1{}^{88}| $, see appendix~\ref{Appendix:B}. Substituting the real fields, $\lambda$ and $\zeta$, in \eqref{A1tens}, we then find  
\begin{equation}\label{superW}
W  \eql \sqrt 2\sqrt{\sinh^6\lambda+\cosh^6\lambda+2 \sinh^3\lambda\cosh^3\lambda\cos(3\zeta)}\,.
\end{equation}
In terms of the superpotential, $W$,  the potential 
\begin{equation}\label{d4Pot}
\cals P\eql -6\cosh(2\lambda)\,,
\end{equation}
is given by 
\begin{equation}\label{}
\cals P\eql {1\over 3}\,\bigg[\left ({\partial W\over\partial\lambda}\right)^2+{4\over \sinh^2(2\lambda)}\,\left ({\partial W\over\partial\zeta}\right)^2\,\Big]-3\,W^2\,.
\end{equation}
Note that unlike the potential, $\cals P$,  the superpotential, $W$, is invariant only under a $\ZZ_3$ subgroup of $\rm U(1)_\zeta$.

\subsection{Domain wall  Ans\"atze and BPS equations}
\label{subsec:domwall}

In this paper we are interested in a special class of solutions corresponding to RG-flows and one-dimensional defects in the dual ABJM theory. Thus we take  the metric 
given by a domain wall Ansatz
\begin{equation}\label{domwAnz}
ds_{1,3}^2\eql e^{2A(r)}ds_{1,2}^2+dr^2\,,
\end{equation}
and where the metric function, $A(r)$, and  the scalar fields, $\lambda(r)$ and $\zeta(r)$, are functions of the radial coordinate, $r$, only.  Furthermore, $ds_{1,2}^2$, is either a Minkowski  metric  (RG-flows) or a metric on $AdS_3$ of radius $\ell$ (Janus solutions),
\begin{equation}\label{ads3}
ds_{1,2}^2\eql e^{2y/\ell}(-dt^2+dx^2)+dy^2\,.
\end{equation}
Since, at least formally, the equations for the RG-flows can be obtained by taking the radius $\ell\to\infty$, throughout much of the discussion we will write only the more general formulae for the Janus solutions.

The equations of motion for the metric \eqref{domwAnz} and the scalar fields that follow from the Lagrangian \eqref{theLag} are 
\begin{equation}\label{d4eqs}
\begin{split}
\lambda'' & \eql -3\, A'\,\lambda'+{1\over 4}\,\sinh(4\lambda)\,(\zeta')^2-2 g^2\,\sinh(2\lambda)\,,\\[6 pt]
\zeta'' & \eql -3\,\,A'\zeta'-4\coth(2\lambda)\,\zeta'\,\lambda'\,,\\[6 pt]
A'' & \eql -{3\over 2}\,(A')^2-{3\over 2}\,(\lambda')^2-{3\over 8}\,\sinh^2(2\lambda)\,(\zeta')^2-{e^{-2A}\over 2\ell^2}\,,\\[6 pt]
\end{split}
\end{equation}
and
\begin{equation}\label{4dEng}
(A')^2-(\lambda')^2-{1\over 4}\sinh(2\lambda)\,(\zeta')^2-2g^2\,\cosh(2\lambda)+{e^{-2A}\over \ell^2}\eql 0\,,
\end{equation}
where last are two equations are independent combinations of the Einstein equations\footnote{As a consequence of the Bianchi identities, the derivative of (\ref{4dEng}) follows from (\ref{d4eqs}).}.

Imposing an unbroken supersymmetry along the flow, one obtains a first order system of the BPS equations. We refer the reader to \cite{Bobev:2013yra} for further details and here only quote the final result:\footnote{Similar BPS equations for holographic domain walls with curved slices were written down in \cite{LopesCardoso:2001rt,LopesCardoso:2002ec,Clark:2005te}.}
\begin{align}
\lambda' & \eql -{1\over 3}\,\bigg({A'\over W}\bigg) \,{\partial W\over\partial\lambda}~+~{2\kappa   \over 3}\, \bigg({e^{-A}\over\ell}\bigg) \,{1\over\sinh(2\lambda) }\,\frac{1}{W}\,{\partial W\over\partial\zeta}\,, \label{JanusBPS:1} \\[10 pt]
\zeta' & \eql-{4\over 3}\,\bigg({A'\over W}\bigg) \,{1 \over\sinh^2(2\lambda) }\,{\partial W\over\partial\zeta} ~-~ {2 \kappa \over 3}\, \bigg({e^{-A}\over\ell}\bigg) \,{1\over\sinh(2\lambda)}\,\frac{1}{W}\,{\partial W\over\partial\lambda}    \,,\label{JanusBPS:2}
\end{align}
together with 
 \begin{equation}\label{Apsq}
(A')^2\eql g^2\,W^2 ~-~{e^{-2A}\over\ell^2} \,.
\end{equation}
The constant $\kappa=\pm 1$ is determined by the chirality of the unbroken supersymmetry, with $\Neql (2,0)$ for $\kappa=1$ and $\Neql (0,2)$ for $\kappa=-1$. In the following we  set $\kappa=1$.
Note that \eqref{Apsq} is the same as \eqref{4dEng} after one eliminates the derivatives of the scalar fields   using \eqref{JanusBPS:1} and \eqref{JanusBPS:2}. It is also straightforward to verify that the equations of motion \eqref{d4eqs} follow from the BPS equations. 

Finally, the BPS equations for supersymmetric RG-flows   
\begin{equation} 
A'  ~=~   \pm g \,W \,, \label{RGA}
\end{equation}
and
\begin{equation}\label{RGflows}
 \lambda'   ~=~   \mp {g \over 3}  \, {\partial W\over\partial \lambda} \,, \qquad 
\zeta' ~=~  \mp{ 4\,g \over 3 \, \sinh^2(2\lambda) }\,{\partial W\over\partial\zeta}   \,,
\end{equation}
are obtained from \eqref{JanusBPS:1}, \eqref{JanusBPS:2} and \eqref{Apsq} by taking the $\ell\to\infty$ limit. There is no constraint on the chirality of the unbroken $\Neql 2$ supersymmetry.

\subsection{Integrating the BPS equations}
\label{subSec:intBPS}

\begin{figure}[t]
\centering
\includegraphics[width=8cm]{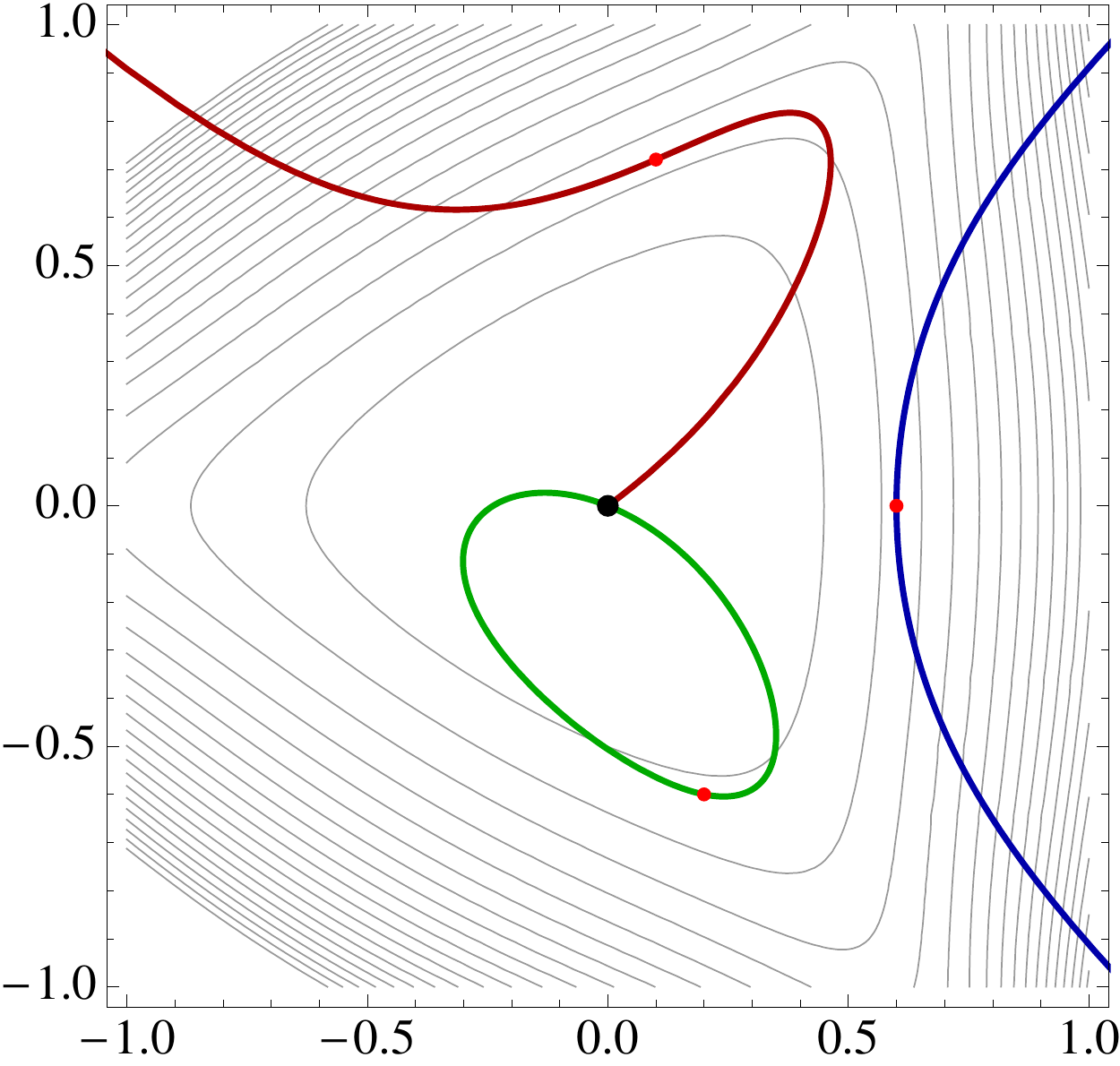}
\caption{Typical flow trajectories for the Janus solutions to  the BPS equations \eqref{JanusBPS:1}--\eqref{Apsq} in the $(\lambda\cos\zeta, \lambda\sin\zeta)$-plane. The background contours are of the  superpotential $W(\lambda,\zeta)$. A red dot denotes the ``central point'' of a flow  at $(\lambda_c\cos\zeta_c,\lambda_c\sin\zeta_c)$ where $A'=0$.}
\label{spagh}
\end{figure}

\begin{figure}[h]
\centering
\includegraphics[width=6. in]{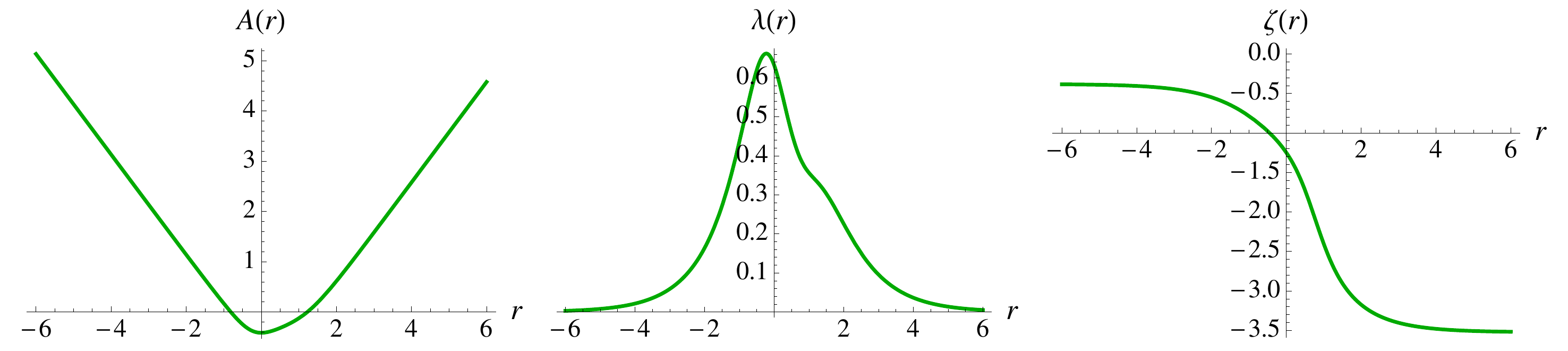}
\\[10 pt]
\includegraphics[width=6. in]{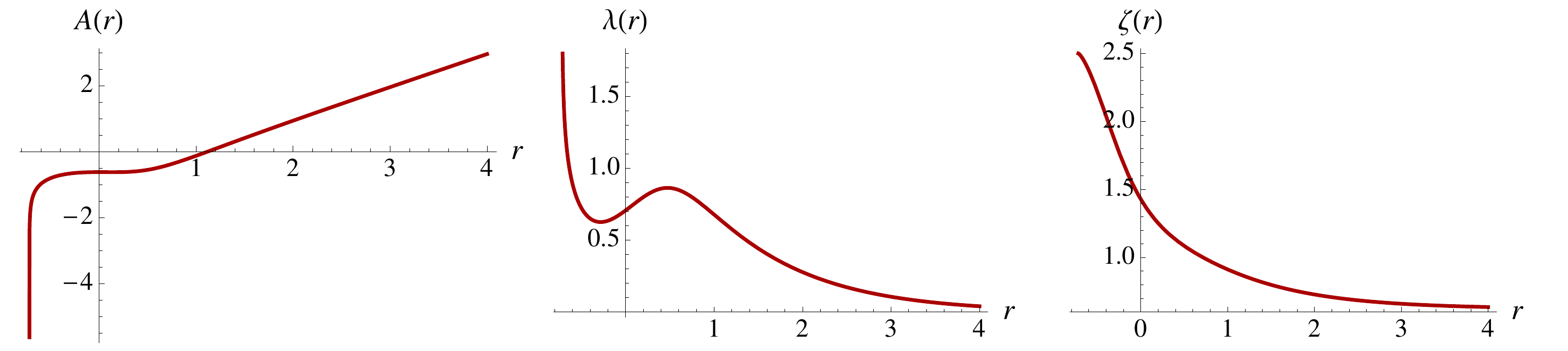}
\\[10 pt]
\includegraphics[width=6. in]{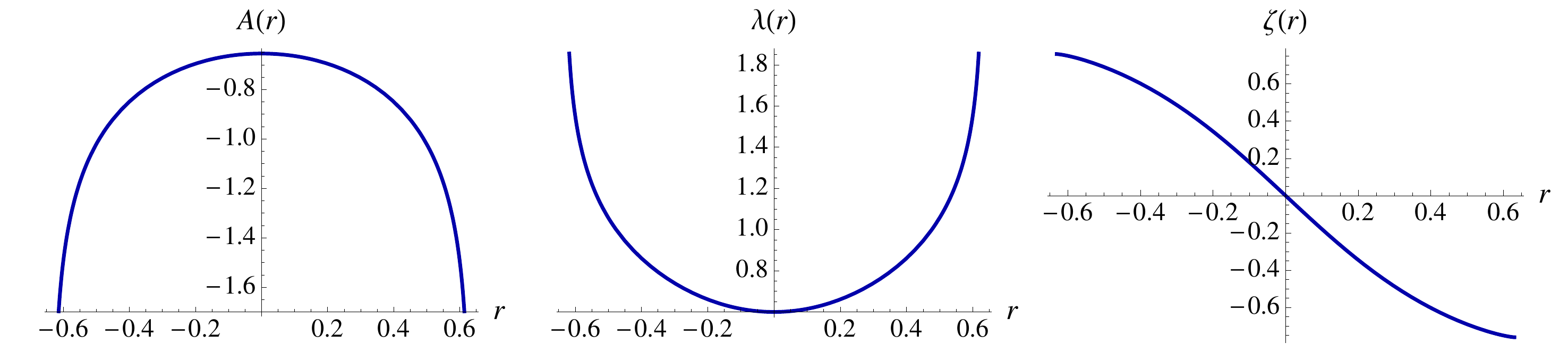}
\caption{Typical  profiles of the metric function, $A(r)$, and the scalar fields, $\lambda(r)$ and $\zeta(r)$, for the different types of   flows in Figure~\ref{spagh}.}
\label{spprofiles}
\end{figure}

\begin{figure}[t]
\centering
\includegraphics[width=8cm]{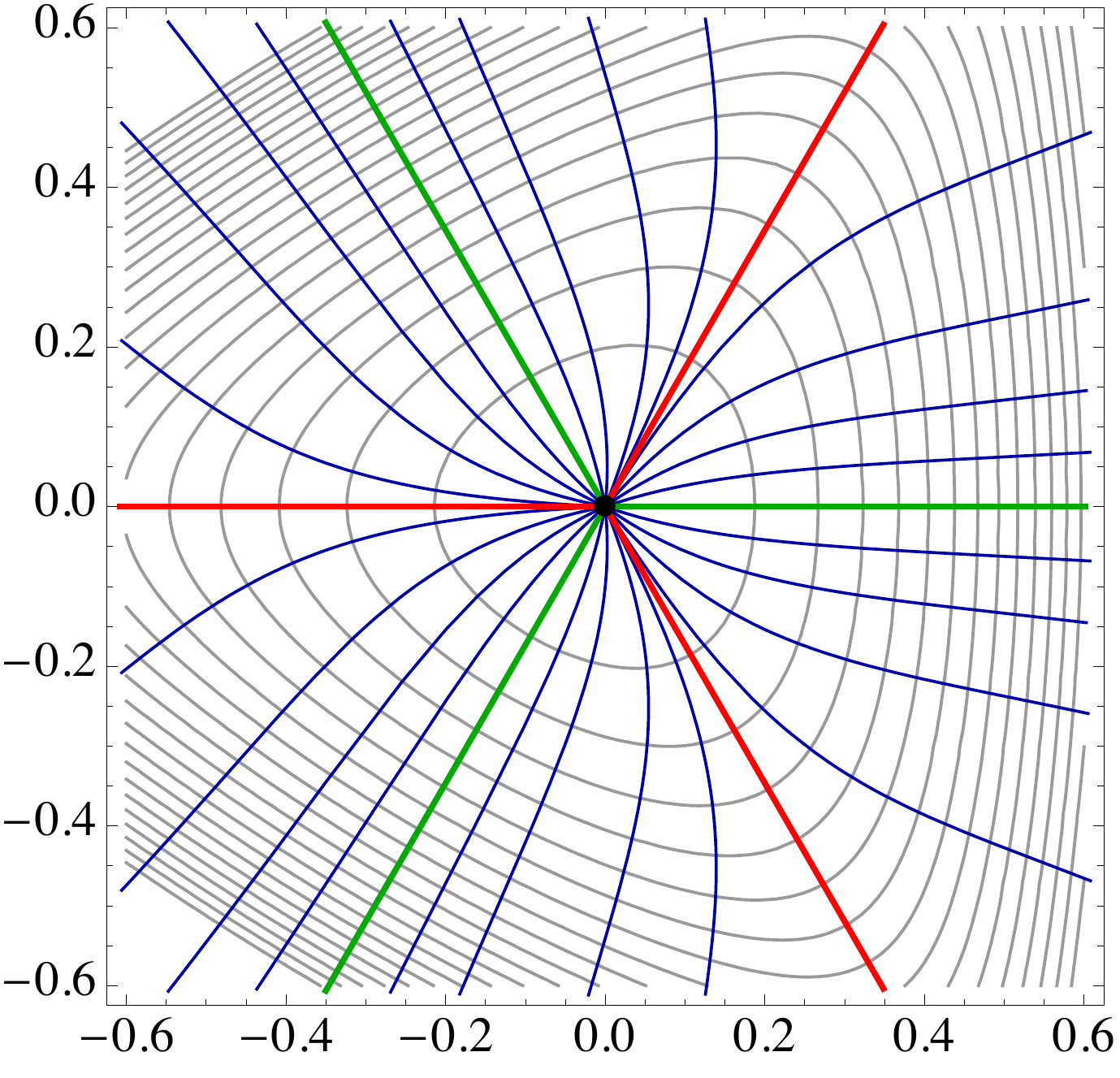}
\caption{RG-flow trajectories in the $(\lambda\cos\zeta, \lambda\sin\zeta)$-plane. The background contours are of the real superpotential $W(\lambda,\zeta)$. The ridge trajectories have constant $\zeta$ with $\cos 3\zeta=1$ (green) and $\cos 3\zeta=-1$ (red), respectively.
}
\label{Rgflows}
\end{figure}

\begin{figure}[t]
\centering
\includegraphics[width=7.5 cm]{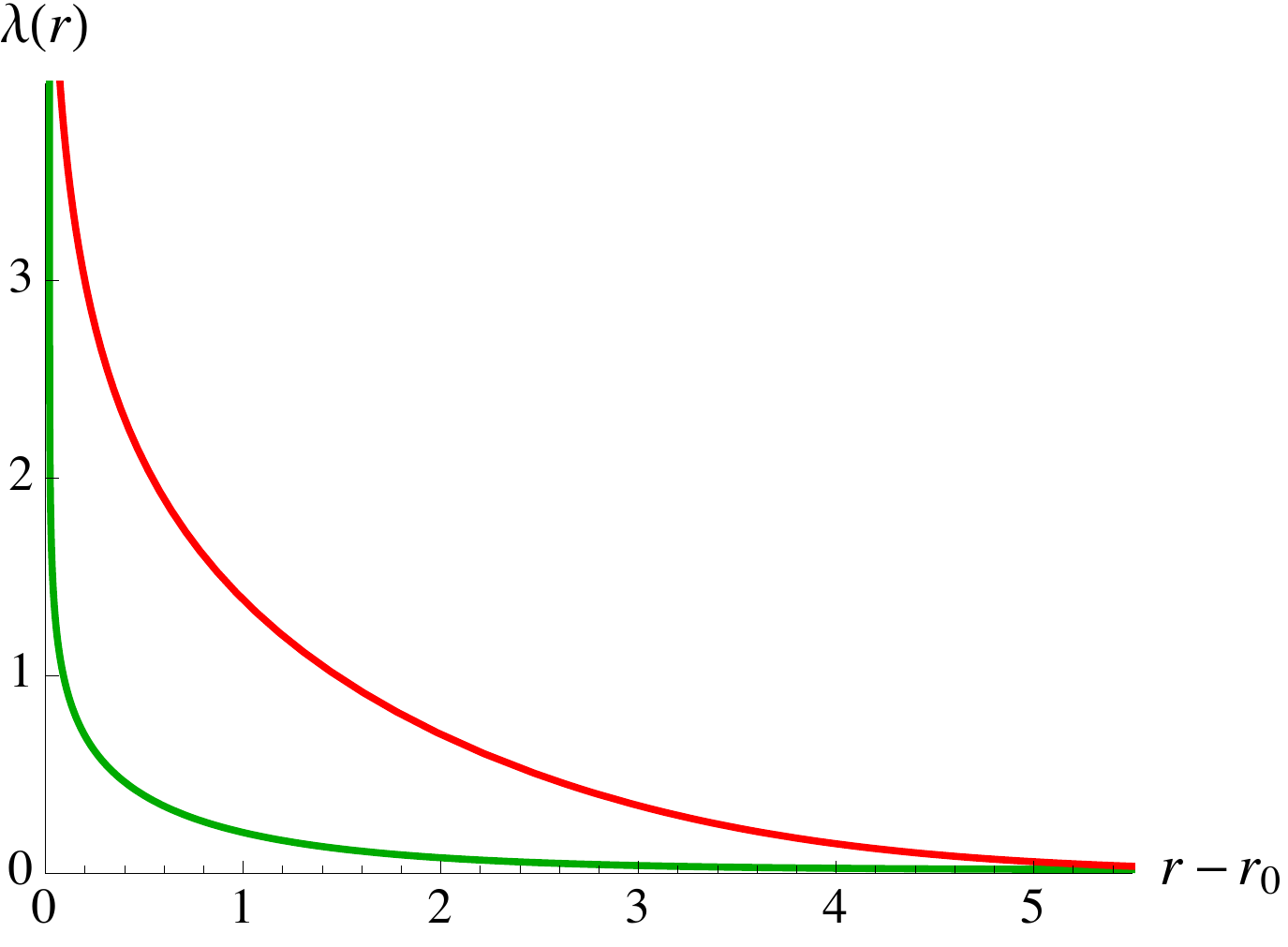}\qquad 
 \includegraphics[width=6 cm]{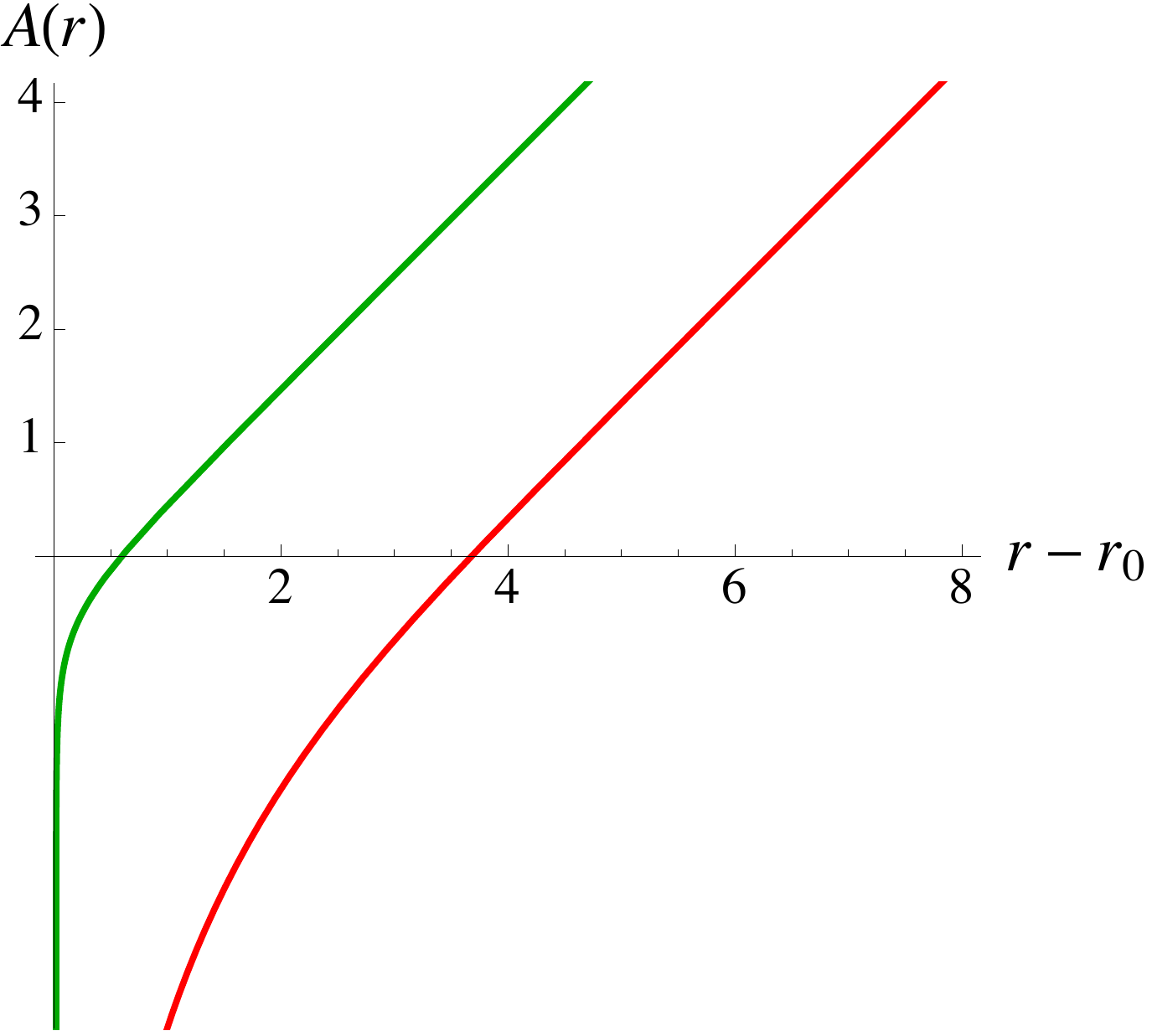}
\caption{Ridge flows for $\cos 3\zeta=-1$ (red) and $\cos 3\zeta=1$ (green) with  $A_0=0$.}
\label{ridgeflows}
\end{figure}

The Janus solutions to the BPS equations \eqref{JanusBPS:1}--\eqref{Apsq}  have been studied in \cite{Bobev:2013yra} where it was shown   by a numerical analysis that there are three classes of  solutions shown in Figures~\ref{spagh} and \ref{spprofiles}: regular Janus solutions (shown in green) interpolating between two $AdS_4$ regions corresponding to the same $\rm SO(8)$ stationary point of the potential \eqref{d4Pot} and singular  solutions that diverge on either one side (shown in red) or both sides (shown in blue) of the flow. The Janus solutions are characterized by the presence of a special central point along a flow   where the solution passes from one branch of \eqref{Apsq} to another. This point is marked by a red dot and the corresponding values of the scalar fields are denoted by $\lambda_c$ and $\zeta_c$, respectively. The position of this point for a given flow determines the type of a solution, see Figure~3 in \cite{Bobev:2013yra}. In particular, for $\cos\zeta_c\not=-1$, all solutions are singular provided  $\lambda_c$ is large enough. It is only when $\cos\zeta_c=-1$ that all solutions are regular Janus solutions irrespective of the value of  $\lambda_c$. 

In addition, there are solutions akin to RG-flows, which asymptote to $AdS_4$ on one side and become singular on the other, while remaining on a single branch of \eqref{Apsq}. They can be thought of as a singular limit of Janus solutions where the central point is moved off to infinity. Simplest examples of such flows are obtained by taking constant $\zeta=\zeta_0$ with $\cos(3\zeta_0)\not=\pm1$. Solving \eqref{JanusBPS:1}--\eqref{Apsq} for $A'$,  $\lambda'$ and $A$, and then imposing consistency between them, one is left with 
\begin{equation}\label{constz}
\begin{split}
\lambda' & \eql \mp{g\over \sqrt 2}\sinh(2\lambda)\sqrt{\cosh(2\lambda)+\cos(3\zeta_0)\sinh(2\lambda)}\,,\\[6 pt]
{e^{-2A}\over\ell^2} & \eql {g^2\over 2}\,{\sin^2(3\zeta_0)\,\sinh(2\lambda)\over \cos(3\zeta_0)+\coth(2\lambda)}\,.
\end{split}
\end{equation}
Choosing the top sign in \eqref{constz}, we can impose the $AdS_4$ boundary condition in the UV, that is $\lambda\to 0$ as $r\to\infty$, to integrate the first equation for $\lambda(r)$, and then   solve the second equation  for $A(r)$. The resulting solutions are similar to the ones in Figure~\ref{ridgeflows}, which we will discuss shortly.

The situation simplifies considerably in the RG-flow limit  where  the scalar equations \eqref{RGflows} do not involve the metric function, $A$, and can be solved first. Choosing the top sign in \eqref{RGA}--\eqref{RGflows}, which corresponds to the UV region at $r\rightarrow\infty$, one then finds flows shown in Figures~\ref{Rgflows} and \ref{ridgeflows}   \cite{Pilch:2015vha}. 

In fact, as we have discussed in \cite{Pilch:2015vha},  generic solutions for the RG-flows can be determined analytically using two constants of   motion: the general one  
\begin{equation}\label{cI1}
\cals I_1 \eql e^{3A}\sinh^2(2\lambda)\,\zeta'\,,
\end{equation}
valid for any $\ell$ and 
corresponding to the conserved current \eqref{consJ}, and the second constant%
\footnote{Given different types of  solutions in Figure~\ref{spagh}, one would not expect to find such an additional constant of motion for the BPS equations \eqref{JanusBPS:1}--\eqref{Apsq} at finite $\ell$.}
\begin{equation}
\begin{split}
 \cI_2 ~=~ &    \frac{W^2}{(\cosh 2\lambda + \cos\zeta \, \sinh 2\lambda )^3  }   \,   \frac{\sin 3\, \zeta}{\sin^3 \zeta } \\[6 pt]
  ~=~ &  \frac{(4 \, \cos^2 \zeta -1)}{2 \,\sin^2 \zeta }\,\frac{(3+(\cosh 2\lambda -2 \cos\zeta \, \sinh 2\lambda )^2) }{(\cosh 2\lambda + \cos\zeta \, \sinh 2\lambda )^2  } \,,
 \end{split} \label{Iom2}
\end{equation}
for the first order system \eqref{RGA}--\eqref{RGflows}. Using $\cals I_1$ and $\cals I_2$, one can then determine $A$ and $\zeta$ as a function of $\lambda$, which is sufficient given the reparametrization invariance for the coordinate along the flow. 

However, this method of integration fails  for the special flows with constant $\zeta=\zeta_0$. For an RG-flow, one must then have  $\cos3\zeta_0=\pm1$ (see, green and red ridge flows in Figure~\ref{Rgflows}). The resulting equations can be obatained from \eqref{constz} by taking the limit,
\begin{equation}\label{}
\cos(3\zeta_0)~\longrightarrow~\pm1\,,\qquad \ell~\longrightarrow~\infty\,,\qquad \ell\,g \,\sin(3\zeta_0)~\longrightarrow~2\sqrt 2\,{e^{-A_0} }\,,
\end{equation}
where $A_0$ is a constant. One can then integrate those  equations directly to obtain 
\begin{equation}\label{sollambda}
\mathop{\rm arccoth}(e^{\lambda })\pm \arctan(e^{\lambda })\mp {\pi\over 2} ~=~ {g\over\sqrt 2}(r-r_0)\,,
\end{equation}
and
\begin{equation}\label{solA}
A(r) ~=~ A_0-\log(e^{4\lambda}-1)+
\begin{cases}
3 \lambda & \text{for $\cos 3\zeta=+1$}\\
\lambda &  \text{for $\cos 3\zeta=-1$}
\end{cases} 
\end{equation}
where $r_0$ and $A_0$ are integration constants. The solutions to those ridge flows are shown in Figure~\ref{ridgeflows}. Comparing with the trajectories in Figure~\ref{Rgflows}, we expect that the solutions with $\cos(3\zeta_0)=1$ are representative of the generic RG-flows, while those with $\cos(3\zeta_0)=-1$ are special. This expectation is confirmed by the asymptotic expansions that we will now discuss. 

\subsection{Behavior at large $\lambda$}
\label{subSec:asymptotics}

The holographic RG flows,  governed by  (\ref{RGA}) and (\ref{RGflows}), have $\lambda \to \infty$ at some finite value of $r$. Generically, such solutions  are dual to a massive flow toward some new infra-red limit.   As was observed in  \cite{Pilch:2015vha}, the RG flow solutions considered here can encode rich and interesting infra-red physics once one examines them  in M theory.    The Janus flows, governed by (\ref{JanusBPS:1})--(\ref{Apsq}),  can either form a loop starting and finishing at  $\lambda=0$ or start at  $\lambda=0$ and ultimately flow with  $\lambda \to \infty$.  There are also Janus solutions that begin and end with $\lambda \to \infty$.  The Janus flows that involve large $\lambda$ may be viewed as interfaces that explore the infra-red structure of the holographic flow solutions.    We  will therefore examine the limiting behaviors of  these flows as  $\lambda \to \infty$.  In Section \ref{Sec:LargeLambda} we will uplift these results to  M theory to see more precisely how they may be interpreted in terms of M branes.

From the explicit solutions in Figures~\ref{spprofiles} and \ref{ridgeflows}, we see that the limit is characterized by  
\begin{equation}\label{}
\lambda~\longrightarrow~ \infty\,,\qquad \zeta~\longrightarrow~ \zeta_\infty\,,\qquad A~\longrightarrow~  -\infty \qquad \text{as}\quad r~\longrightarrow~  r_0\,,
\end{equation}
where $\zeta_\infty$ is a constant asymptotic angle for a given flow.  This observation is confirmed by a more careful expansion  of the BPS equations \eqref{JanusBPS:1}--\eqref{RGflows}.

Expanding the superpotential \eqref{superW} as  $\lambda\to\infty$ for a generic $\zeta$, we have
\begin{equation}\label{genexpW}
W\til {1\over 4}\sqrt{1+\cos(3\zeta)}\,e^{3\lambda}+O(e^{-\lambda})\,,\qquad \cos(3\zeta)~\not=~-1\,,
\end{equation}
while for the special  ridge flows,
\begin{equation}\label{}
W\til {3\over 2\sqrt 2}\,e^\lambda+O(e^{-3\lambda})\,,\qquad \cos(3\zeta)\eql-1\,.
\end{equation}
The flow equations  (\ref{RGflows}) and  (\ref{RGA}) implify
\begin{align}
{dA\over d\lambda} &~=~  -3\, W \bigg({\partial W\over\partial\lambda}\bigg)^{-1} ~\sim~\begin{cases}
-3  & \text{for $\cos 3\zeta= -1$}  \\
-1 &  \text{otherwise}
\end{cases}   \label{dAdlambda}\,. \\[10 pt]
{d \zeta \over d\lambda} & ~=~  { 4 \over  \sinh^2(2\lambda) }\,{\partial W\over\partial\zeta}\, \bigg({\partial W\over\partial\lambda}\bigg)^{-1}  \quad \to \quad 0  \,.\label{dzetadlambda}
\end{align}
The latter confirms the constancy of $\zeta$ at infinity while the former shows the rate of divergence of $A$ depends upon that angle.  This will translate into different physics once we uplift those flows to M theory.

\section{The uplift}
\label{Sect:uplift}

We have obtained the  Lagrangian \eqref{theLag} and the BPS equations \eqref{JanusBPS:1}--\eqref{RGflows}
by a consistent truncation of the bosonic Lagrangian and the supersymmetry variations of four-dimensional, $\Neql8$  gauged supergravity. Since the latter theory is a consistent truncation of M theory on $S^7$ \cite{deWit:1986mz,de Wit:1986iy,Nicolai:2011cy}, any solution of the equations of motion for the Lagrangian \eqref{theLag}  can be uplifted to a solution of the eleven-dimensional supergravity. In the next two sections we verify this   using   explicit uplift formula for the metric \cite{deWit:1984nz} and the recently obtained uplift formulae for the flux \cite{deWit:2013ija,Godazgar:2013nma,Godazgar:2013dma,Godazgar:2013pfa,Godazgar:2014nqa,Godazgar:2015qia}.   

Similar calculation verifying the new uplift Ans\"atze for the flux has been carried out recently for a special class of solutions of the four-dimensional, $\Neql8$  theory given by some of the   stationary points of the potential: $\rm SO(8)$, $\rm SO(7)^\pm$, $\rm G_2$, $\rm SU(4)^-$ \cite{Godazgar:2013nma,Godazgar:2013pfa}, and SO(3)$\times$SO(3) \cite{Godazgar:2014eza}, for which the four-dimensional space-time is $AdS_4$ and the scalar fields  are constant. Our construction of the uplift is similar as in those references, which the reader should consult for any omitted background material.

\subsection{$\rm SU(3)$$\times$$\rm U(1)$$\times$$\rm U(1)$ invariants on $S^7$}
\label{subSec:invariants}

The construction of an uplift inevitably leads to rather complicated formulae. Both to organize the calculation and to write down the result in a sucinct form, it is convenient to express the internal components of the fields in terms of canonical  $\rm SO(7)$  tensors on $S^7$ that are associated with the $\rm E_{7(7)}$ generators of the scalar fields in the truncation. To this end let us define:\footnote{For a more extensive discussion of these tensors, see  \cite{Godazgar:2014eza} and the original references therein.} 
\begin{equation}\label{thexis}
\xi_{mn}\eql -{1\over 16}\Phi^+_{IJKL}K_m^{IJ}K_n^{KL}\,,\qquad 
\xi_m\eql {1\over 16}\Phi^+_{IJKL}K_{mn}^{IJ}K^{n\,KL}\,,\qquad 
\xi\eql \go^{mn}\xi_{mn}\,,
\end{equation}
and
\begin{equation}\label{theS}
S_{mnp}\eql {1\over 16} \Phi^-_{IJKL} K_{mn}^{IJ}K_p^{KL}\,,
\end{equation} 
where $\Phi^\pm_{IJKL}$ are the   $\rm SO(8)$ tensors defined in \eqref{Ytens} and
\begin{equation}\label{Killvec}
 K^m{}^{IJ} = i \bar \eta^I \Gao^m \eta^J\,, \qquad
  K_{mn}^{IJ} = \bar \eta^I \Gao_{mn} \eta^J\,,\qquad \Do_m K_n^{IJ}\eql m_7\,K_{mn}^{IJ}\,,
  \end{equation}
are the $\rm SO(8)$ Killing vectors (one-forms) and two-forms on the round $S^7$ given in terms of  an orthonormal basis of Killing spinors, $\eta^I$, 
\begin{equation}\label{}
 \label{killspineqn}
  i \Do_m \eta^I = \frac{m_7}{2} \Gao_m
  \eta^I\,,\qquad \bar\eta^I\eta^J\eql\delta^{IJ}\,,\qquad I,J=1,\ldots,8\,.
\end{equation}
The inverse  radius of $S^7$ is denoted by  $m_7\equiv L^{-1}$ and   $\Gamma^m=\eo_a{}^m\Gamma^a$. The circle indicates that $\eo^a$ is a siebenbein for the round metric on $S^7$, $ds_{S^7}^2=\go_{mn}dy^mdy^n$ where $\go_{mn}=\eo_m{}^a\eo_n{}^b\delta_{ab}$,  and $\Do_m$ is the covariant derivative with respect to that metric. 
Unless indicated otherwise, all indices on the $S^7$ tensors are raised and lowered with the {\it round} metric, for example $K_m^{IJ}=\go_{mn}K^{nIJ}$. The coordinates, $y^m$, on $S^7$ are for the moment arbitrary. However, one should note that $\xi$ defined in \eqref{thexis} is a scalar harmonic on $S^7$ and may be thought of as providing a natural internal coordinate on the compactification manifold.

By construction, the tensors \eqref{thexis} and \eqref{theS} are invariant, {\it i.e.} have vanishing Lie derivative, under the \suthuu$\,\subset \rm SO(8)$ symmetry group of the truncation.   In particular, the Killing vectors for the two $\rm U(1)$'s, 
\begin{equation}\label{}
\omg_m\eql \Omega_{IJ}^\alpha K_m^{IJ}\,,\qquad w_m\eql \Omega_{IJ}^\beta K_m^{IJ}\,,
\end{equation}
\begin{equation}\label{}
\Omega_{12}^\alpha \eql \Omega_{34}^\alpha\eql \Omega_{56}^\alpha\eql 1\,,\qquad \Omega^\beta_{78}\eql 1\,,
\end{equation}provide additional invariant one-forms on $S^7$.  In the following we will show that   the metric  and the flux for the uplift can be simply written  in terms of the round metric, $\go_{mn}$, the one-forms 
\begin{equation}\label{}
\xi_{(1)}~\equiv~\xi_m dy^m\,,\qquad \omg_{(1)}~\equiv~\omg_m dy^m\,,\qquad \varth_{(1)}\eql\varth_m dy^m\,,
\end{equation}
the three-form, 
\begin{equation}\label{}
S_{(3)}~\equiv~{1\over 6}S_{mnp}dy^m\wedge dy^n\wedge dy^p\,,
\end{equation}
 and the scalar, $\xi$.

\subsection{The metric}
\label{subSec:metric}

The eleven-dimensional space-time   for  the uplifted solutions  is a warped product, $\cals M_{1,3}\times \cals M_{7}$, with the metric
\begin{equation}\label{d11met}
ds^2_{11}\eql \Delta^{-1}ds_{1,3}^2+ds_7^2\,,
\end{equation}
where\footnote{To distinguish the components of the four-dimensional metric \eqref{domwAnz} from the components of its eleven-dimensional counterpart along the four dimensions, we will denote the former by $\go_{\mu\nu}$. Thus $g_{\mu\nu}=\Delta^{-1}\go_{\mu\nu}$.}
 $ds_{1,3}^2=\go_{\mu\nu}dx^\mu dx^\nu$ is the metric in four dimensions for a particular solution at hand. The internal metric, $ds_7^2=g_{mn}dy^mdy^n$, is determined  by the celebrated formula for its densitized inverse \cite{deWit:1984nz}:
\begin{equation} \label{ans:inmetric}
\Delta^{-1} g^{mn}  = \frac{1}{8} K^{m\,IJ}  K^{n\,KL} 
   \Big[\left( u^{MN}{}_{IJ} + v^{MNIJ} \right) 
    \left( u_{MN}{}^{KL} + v_{MNKL} \right) \Big]\,,  
\end{equation}
from which the warp factor, $\Delta$,  can be   calculated using
\begin{equation}\label{warpeqs}
\Delta^{-9}\eql \det(\Delta^{-1}g^{mn}\go_{np})\,.
\end{equation}

While it is possible to express the densitized metric entirely using   tensors \eqref{thexis} and \eqref{theS} and their (contracted) products,\footnote{See, for example a general discussion in \cite{Godazgar:2014eza}.} the simplest expression is obtained by noting that the symmetric tensors resulting from such contractions can be rewritten using the round metric, $\go_{mn}$, and  bilinears in the one forms $\xi_m$, $\omg_m$ and $\varth_m$, as in the following examples:
\begin{equation}\label{ident}
\begin{split}
\xi_{mn} 
& \eql {1\over 6}(3+\xi)\go_{mn}+{1\over 6(\xi-3)}\,\xi_m\xi_n+{3\over 8(\xi-3)}\,(\omg_m+\varth_m)(\omg_n+\varth_n)\,,\\[6 pt]
S_{mpq}S_n{}^{pq}& \eql {1\over 4}(\omg_m-\varth_m)(\omg_n-\varth_n)\,,\\[6 pt]
S_{mpr}S_{nq}{}^{r}\xi^p\xi^q& \eql {9\over 4} \,\omg_m\omg_n+ {3\over 4}(9-2\xi)\,\varth_m\varth_n+{3\xi\over 2} \,\omg_{(m}\varth_{n)}\,.
\end{split}
\end{equation}
After some algebra, we then find
\begin{equation}\label{ingginv}
\Delta^{-1} g^{mn} \eql c_1\, \go^{\,mn}+c_2\, \xi^m\xi^n + c_3 \, \omg^m\omg^n+ c_4 \,\varth^m\varth^n + c_5 \,\omg^{(m}\varth^{n)}\,,
\end{equation}
where all the $c_i$ can be expressed in terms of four-dimensional quantities and the scalar, $\xi$:
\begin{equation*}\label{}
\begin{split}
c_1 & \eql \cosh (2 \lambda )-\frac{1}{6} (\xi +3) \sinh (2 \lambda ) \cos (\zeta )\,,\\[6 pt]
c_2 & \eql \frac{1}{ 6 (3- \xi )}\,\sinh (2 \lambda ) \cos (\zeta )\,,\\[6 pt]
\end{split}
\end{equation*}
\begin{equation}\label{}
\begin{split}
c_3 & \eql {\sinh(2\lambda)\over 32(\xi-3)}\Big[(\xi -3) \cosh (4 \lambda ) \cos (\zeta )+(\xi -3) \sinh (4 \lambda )-(\xi +9) \cos \zeta \Big]\,,\\[6 pt]
c_4 & \eql{\sinh(2\lambda)\over 16(\xi-3)} \Big[(\xi -3) \sinh ^2(2 \lambda ) \cos (3 \zeta )+\frac{1}{2} (\xi -3) \sinh (4 \lambda )-6 \cos
   \zeta \Big]\,,\\[6 pt]
c_5 & \eql {\sinh(2\lambda)\over 16(\xi-3)} \Big[ (\xi -3) \sinh (4 \lambda ) \cos (2 \zeta )+(\xi -3) \cosh (4 \lambda ) \cos \zeta -(\xi +9)
   \cos \zeta \Big]\,.
\end{split}
\end{equation}
Note that the indices on the right hand side in \eqref{ident} are raised using the round metric, $\go_{mn}$, which is the convention followed throughout this section.

All that is needed now to invert the densitized metric \eqref{ingginv} are   contraction identities between the one-forms, which can be derived   using the explicit form of the $\rm SO(8)$ tensors and properties of the Killing vectors summarized in   \cite{Godazgar:2013nma,Godazgar:2014eza} and the references therein. We have
\begin{equation}\label{contvec}\begin{split}
\xi^m\xi_m   \eql 27- & 6\,\xi-\xi^2\,,\qquad \omg^m\omg^m\eql 12-{8\over 3}\,\xi\,,\qquad \varth^m\varth_m\eql 4\,,\\
& \xi^m\omg_m\eql \xi^m\varth_m\eql 0\,,\qquad \omg^m\varth_m\eql -{4\over 3}\,\xi\,.
\end{split}
\end{equation}
It is then straightforward to check that
\begin{equation}\label{thedenmetric}
\Delta g_{mn}\eql g_1\, \go_{mn}+g_2\, \xi_m\xi_n + g_3 \, (\omg_m\omg_n+ \varth_m\varth_n) + g_4 \,\omg_{(m}\varth_{n)}\,,
\end{equation}
where
\begin{equation}\label{}
\begin{split}
g_1 & \eql  {6\over 6 \cosh (2 \lambda )-(\xi +3) \sinh (2 \lambda ) \cos \zeta }\,,\\[6 pt]
g_2 & \eql {\cals D \over 36(\xi-3)}\,\cos \zeta \big[3 \sinh (4 \lambda )-(\xi +3) \sinh ^2(2 \lambda ) \cos \zeta\big]\,,\\[6 pt]
g_3 & \eql {\cals D\over 16(\xi-3)}\,\big[3 \sinh (4 \lambda ) \cos (\zeta )-\sinh ^2(2 \lambda ) (\xi +3 \cos (2 \zeta ))\big]\,,\\[6 pt]
g_4 & \eql  {\cals D\over 8(\xi-3)}\,\big[3 \sinh (4 \lambda ) \cos (\zeta )-\sinh ^2(2 \lambda ) (\xi  \cos (2 \zeta )+3)\big]\,,
\end{split}
\end{equation}
and
\begin{equation}\label{theDD}
\cals D \eql {36\over \big[\sinh (2 \lambda ) \cos  \zeta  +\cosh (2 \lambda )\big] 
\big[6\cosh (2 \lambda )-(\xi +3) \sinh (2 \lambda ) \cos \zeta \big]^{2}}\,.
\end{equation}
Using the contractions \eqref{contvec}, one can also calculate the derivatives of
the warp factor given by \eqref{warpeqs}
with respect to $\lambda$ and $\zeta$. Then a simple integration yields
\begin{equation}\label{warpDel}
\Delta\eql \cals D^{1/3}\,,
\end{equation}
with the  overall normalization set by $\Delta=1$ for the round sphere metric when $\lambda=0$. Dividing out $\Delta$ in \eqref{thedenmetric}, yields the internal metric, $g_{mn}$, in terms of the $\rm SO(7)$ tensors associated with the truncation.

\subsection{Internal coordinates and local expressions}
\label{subSec:coordinates}

In addition to the metric tensor, we will  also need  the corresponding orthonormal frames and those turn out to be rather cumbersome to write down using the invariant tensors \eqref{thedenmetric}. Also, the formulae like \eqref{thedenmetric} tend to obscure the underlying geometry of the solution and its symmetry. To address both of these points, we will now choose a suitable set of coordinates, $y^m$, on the internal manifold. As usual, see for example  \cite{Godazgar:2014eza} section 7.1, this can be done systematically as follows: 

First, we embed  $S^7$ into the ambient $\RR^8$  as the surface 
\begin{equation}\label{thesphere}
Y^A Y^A\eql m_7^{-1}\,,
\end{equation}
such that the Killing vectors $K^{IJ}=K_m^{IJ}dy^m$ defined in \eqref{Killvec} are related by triality to the  familiar ones, that is
\begin{equation}\label{triality}
K^{IJ}\eql -{m_7\over 2}\Gamma_{AB}^{IJ} \cals K^{AB}\,,\qquad \cals K^{AB}\eql -{1\over 8m_7} \Gamma_{AB}^{IJ} K^{IJ}\,,
\end{equation}
where
\begin{equation}\label{}
\cals K^{AB}\eql Y^AdY^B-Y^BdY^A\,,
\end{equation}
Similarly, we have
\begin{equation}\label{K2form}
K_{(2)}^{IJ}\equiv {1\over 2}\,K_{mn}^{IJ}dy^m\wedge dy^n\eql {1\over 2}\,\Gamma_{AB}^{IJ}d\cals K^{AB}\,.
\end{equation}

The action of the symmetry group \suthuu\  in the ambient space is given by the branching\footnote{We follow here the usual convention that the supersymmetries, $\epsilon^i$, transform in $\bfs 8_v$, while the ambient coordinates, $Y^A$, in $\bfs 8_s$ of $\rm SO(8)$.}
\begin{equation}\label{8sbranch}
\bfs 8_{s}\quad \rightarrow \quad (\bfs 3,-\coeff{1}{2},\coeff{1}{2})+(\bar{\bfs 3},\coeff{1}{2},-\coeff{1}{2})+(\bfs 1,\coeff{3}{2},\coeff 1 2)+(\bfs 1,-\coeff 3 2,-\coeff 1 2)\,.
\end{equation}
One can choose a representation of   $\Gamma$-matrices  such that the $\rm SU(3)$ generators  act in the subspace, $Y^1,\ldots,Y^6$, while the two $\rm U(1)$ generators  have $2\times 2$ diagonal blocks. Then
a convenient choice for the coordinates, $(y^m)=(\chi,\theta,\alpha_1,\alpha_2,\alpha_3,\psi,\phi)$, on $S^7$, that makes the symmetry manifest is as follows:
\begin{equation}\label{intcoor}
\begin{split}
Y^1+i\, Y^2 & \eql m_7^{-1}\,\cos\chi\sin\theta\,\sin{\frac{\alpha_1}{2}}\,e^{{i\over 2}(\alpha_2-\alpha_3)} e^{-i(\phi+\psi)}\,,\\[6 pt]
Y^3+i\, Y^4 & \eql m_7^{-1}\,\cos\chi\sin\theta\,\cos{\frac{\alpha_1}{2}}\,e^{-{i\over 2}(\alpha_2+\alpha_3)} e^{-i(\phi+\psi)}\,,\\[6 pt]
Y^5+i\, Y^6 & \eql m_7^{-1}\cos\chi\cos\theta \,e^{-i(\phi+\psi)} \,,\\[6 pt]
Y^7+i\, Y^8 & \eql m_7^{-1} \sin\chi\,e^{-i\phi} \,,\\ 
\end{split}
\end{equation}
where $\alpha_1,\alpha_2,\alpha_3$ are the $\rm SU(2)$ Euler angles, while the angles $\psi$ and $\phi$ parametrize the U(1)$\times$U(1) isometry.\footnote{More precisely, the two U(1) angles are $\phi+\psi/2$ and $-\phi-3\psi/2 $, respectively.}  In this parametrization, the round metric on $S^7$ with unit radius is\footnote{All  functions and forms in the ambient $\RR^8$ are implicitly pulled-back onto $S^7$ using \eqref{thesphere}.}
\begin{equation}\label{s7metr}
\begin{split}
ds_{S^7}^2 & ~\equiv~ m_7^2\,dY^AdY^A\\
& \eql d\chi^2+\cos^2\chi\,\Big[ds_{\CC\PP^2}^2+\sin^2\chi\big(d\psi+{1\over 2}\sin^2\theta\,\sigma_3\big)^2\Big]\\[6 pt]
& \hspace{42 pt} +\Big[d\phi+\cos^2\chi\,\big(d\psi+{1\over 2}\sin^2\theta\,\sigma_3\big)\Big]^2\,,
\end{split}
\end{equation}
where
\begin{equation}\label{}
ds_{\CC\PP^2}^2\eql d\theta^2+{1\over 4}\sin^2\theta\big(\sigma_1^2+\sigma_2^2+\cos^2\theta \sigma_3^2)\,,
\end{equation}
is the metric on $\CC\PP^2$ and $\sigma_i$ are the $\rm SU(2)$-invariant forms. The first line in \eqref{s7metr} is the metric on $\CC\PP^3$ and the second line is the Hopf fiber. The $\rm SU(3)\times U(1)_\psi$ symmetry acts transitivly on $\CC\PP_2$ and the $\psi$-fiber. Both the metric on $\CC\PP^2$ and the one form    $d\psi+{1\over 2}\sin^2\theta$  are invariant.

Using \eqref{triality} and \eqref{intcoor}, we can now express the invariants introduced in section~\ref{subSec:invariants} in terms of ambient and local coordinates. We find that the invariant function, $\xi$, is simply
\begin{equation}\label{xifunct}
\begin{split}
\xi & \eql 3-12\,m_7^2\big[(Y^7)^2+(Y^8)^2\big]\eql -9+12\,m_7^2\big[(Y^1)^2+\ldots+(Y^6)^2\big]\\[6 pt]
& \eql 3(1-4 \sin^2\chi)\,,
\end{split}
\end{equation}
while the invariant one-forms are
\begin{equation}\label{oneforms}
\begin{split}
\xi_{(1)} & \eql -12\,m_7(Y^7dY^7+Y^8dY^8)\eql -6\,m_7^{-1}\,\sin(2\chi)\,d\chi\,,\\[6 pt]
\omg_{(1)}+\varth_{(1)} & \eql 8\, m_7(Y^7dY^8-Y^8dY^7)\eql -8\,m_7^{-1}\sin^2\chi\,d\phi\,,\\[6 pt]
\omg_{(1)}-3\varth_{(1)} & \eql 8\, m_7(Y^1dY^2-Y^2dY^1+\ldots+Y^5dY^6-Y^6dY^5)\\
& \qquad \eql
-8 m_7^{-1}\cos^2\chi\,(d\phi+d\psi+{1\over 2}\sin^2\theta\sigma_3)\,.
\end{split}
\end{equation}
We also have 
\begin{equation}\label{S3loc}
\begin{split}
S_{(3)}\equiv{1\over 6}\,S_{mnp}\,dy^m\wedge dy^n\wedge dy^p & \eql -{1\over 6}\,m_7\,\Phi^+_{MNPQ}\,Y^M\,dY^N\wedge dY^P\wedge dY^Q\\[6 pt]
& \eql -m_7^{-3}\,J_{\CC\PP^3}\wedge \vartheta_{S^7}\,,
\end{split}
\end{equation}
where 
\begin{equation}\label{}
J_{\CC\PP^3}\eql {1\over 2}\,d\vartheta_{S^7}\,,\qquad \vartheta_{S^7}\eql d\phi+\cos^2\chi\,(d\psi+{1\over 2}\sin^2\theta\,\sigma_3)\,,
\end{equation}
are, respectively, the complex structure on $\CC\PP^3$ and the corresponding Sasaki-Einstein one-form on $S^7$.

Finally, we substitute the invariants \eqref{xifunct} and \eqref{oneforms} into the warp factor \eqref{warpDel} and the metric \eqref{thedenmetric}.  To simplify expressions we define: 
\begin{equation}
 \DX_\pm(x)  ~\equiv~  \cosh 2\lambda \pm \cos \zeta  \,\sinh 2\lambda\,, \qquad  \Sigma(x,\chi)  ~\equiv~   \DX_+\, \sin^2\chi +    \DX_- \,\cos^2\chi  \,,\label{defns1}
\end{equation}
which are functions   of the space-time coordinates, $x^\mu$, and the internal coordinate, $\chi$. Then the internal metric can be written as
\begin{equation}\label{theintmet}
\begin{split}
ds_7^2 & \eql m_7^{-2}\Big({\Sigma\over \DX}\Big)^{2\over 3}\Big[\,d\chi^2+\cos^2\chi\,{\DX\over\Sigma}\Big( \,ds_{\CC\PP^2}^2+\sin^2\chi\, {\DX\over\Sigma}\,(d\psi+{1\over 2}\sin^2\theta\,\sigma_3+{\Xi\over \DX}\,d\phi)^2\Big)\\[6 pt]
&\hspace{85 pt}+ {1\over\Sigma^2}\,\Big(d\phi+\cos^2\chi\,\big(d\psi+{1\over 2}\sin^2\theta\,\sigma_3\big)\Big)^2\,\Big]\,,
\end{split}
\end{equation}
where to simplify the notation we  set $\DX\equiv \DX_+$ and $\Xi\equiv \DX_+-\DX_-$. The warp factor \eqref{warpDel} is  
\begin{equation}\label{}
\Delta={1\over X^{1/3}\Sigma^{2/3}}\,.
\end{equation}

For $\lambda=0$, we have $\DX_\pm=\Sigma=1$ and the metric \eqref{theintmet} reduces to the metric \eqref{s7metr} on the sphere with radius $m_7^{-1}$. The deformation clearly preserves the \suthuu\ symmetry as well as the metric along the Hopf fiber, which is now rescaled by $\Sigma^{-2}$ with respect to the six-dimensional base. This suggests that there might be some deformed K\"ahler geometry still present in the background. We will return to this point below in section~\ref{Sec:LargeLambda}.  

\subsection{The transverse flux}
\label{subSec:trflux}

It is rather remarkable that it took more than 25 years to obtain workable formulae for the   four-form flux, $F_{(4)}=dA_{(3)}$. Indeed, while the general proof of the consistent truncation of eleven-dimensional supergravity on $S^7$ \cite{deWit:1986mz,deWit:1986iy,Nicolai:2011cy} yielded   explicit formulae for $F_{(4)}$, those formulae were rather difficult if not impossible to use for all but  the simplest stationary point solutions \cite{Nicolai:2011cy}. It is only recently that  new Ans\"atze for various components of the four-form flux were found in \cite{deWit:2013ija,Godazgar:2013nma,Godazgar:2013dma,Godazgar:2013pfa,Godazgar:2014nqa,Godazgar:2015qia} whose complexity is comparable to  that of the   metric Ansatz.

Starting with a domain wall solution in four-dimensions with a  metric as in \eqref{domwAnz}  and scalar fields depending only on the transverse coordinate, the corresponding four-form flux in eleven-dimensional supergravity  can be decomposed into a sum of two terms
\begin{equation}\label{decflux}
F_{(4)}\eql F_{(4)}^{\rm st}+F_{(4)}^{\rm tr}\,,
\end{equation}
where $F_{(4)}^{\rm st}=F_{(4,0)}+F_{(3,1)}$ is the ``space-time'' flux and $F_{(4)}^{\rm tr}=F_{(0,4)}+F_{(1,3)}$ is the ``transverse'' flux. A label $(4-p,p)$ indicates a $(4-p)^{\rm th}$  order form along $\cals M_{1,3}$ and a $p^{\rm th}$ order form along the internal manifold, $\cals M_7$. Since by the Poincar\'e or conformal symmetry along the three-dimensional slices in $\cals M_{1,3}$ there can be no  $(2,2)$-form in \eqref{decflux},\footnote{Such terms must also vanish whenever the vector fields  in four dimensions are set to zero  \cite{Godazgar:2015qia}.}  the Bianchi identity, $dF_{(4)}=0$, implies that both   $F_{(4)}^{\rm st}$ and $F_{(4)}^{\rm tr}$ must be closed. Hence 
 $F_{(4)}^{\rm tr}=dA_{(3)}^{\rm tr}$, where $A_{(3)}^{\rm tr}$ can have at most   one ``leg'' along $dr$ and thus 
 can be  always  gauge transformed into a 3-form with all three legs along the internal manifold $\cals M_7$, that is      $A_{(3)}^{\rm tr}={1\over 6}A_{mnp}dy^m\wedge dy^n\wedge dy^p$. 

The components $A_{mnp}$ are given by the new uplift Ansatz  \cite{deWit:2013ija,Godazgar:2013nma}, which, in our conventions,  reads  
\begin{equation}\label{our:flux}
 \Delta^{-1}g^{pq} A_{mnp}  
=  {i\over 16}  K_{mn}^{IJ}  K^{q\, KL}  
  \Big[ \left( u^{MN}{}_{IJ} -  v^{MNIJ} \right) \left( u_{MN}{}^{KL} + v_{MNKL} \right) \Big]\,. 
\end{equation}
It is convenient to define a two-form $S_m\equiv {1\over 2}S_{mnp}dy^n\wedge dy^p$.
 Evaluating \eqref{our:flux} in terms of invariants, we then find  
\begin{equation}\label{}
\begin{split}
{1\over 2} \Delta^{-1}g^{pq} A_{mnp}\, dy^m\wedge dy^n & \eql (a_{11}\omg^q+a_{12}\,\varth^q)\,d\omg+
 (a_{21}\omg^q+a_{22}\,\varth^q)\,d\varth + a_3\,S^q\,,
\end{split}
\end{equation}
where the vector index on the right hand side is raised using the round metric and the coefficients are given by \begin{equation}\label{}
\begin{split}
a_{11}& \eql  {1\over 64}\,m_7^{-1}\sinh^3(2\lambda )\sin\zeta  \,,\\[6 pt]
 a_{12} & \eql \frac{1}{64}\, m_7^{-1}\sinh ^2(2 \lambda  ) \sin\zeta\,\big[2\cosh(2\lambda)\cos\zeta -\sinh(2\lambda) \big]\,,\\[6 pt]
a_{21} & \eql  - \frac{1}{64}\, m_7^{-1}\sinh ^2(2 \lambda  ) \sin\zeta\,\big[2\cosh(2\lambda)\cos\zeta +\sinh(2\lambda) \big] \,,\\
   a_{22}& \eql -{1\over 64}\, m_7^{-1}\sin(3\zeta )\sinh^3(2\lambda )\,,\\[6 pt]
a_3 & \eql -{1\over 2}\sin\zeta \sinh(2\lambda )\,.
\end{split}
\end{equation}
Contracting with the densitized metric \eqref{thedenmetric} and then using the 
contraction identities \eqref{contvec} together with 
\begin{equation}\label{contrS}
\begin{split}
\xi^m S_m  & \eql -{3\over 4}\,\omg_{(1)}\wedge\varth_{(1)}\,,\\[6 pt]
\omg^m S_m & \eql {1\over 12}\,m_7^{-1}(\xi-6)\,d\omg_{(1)}+{1\over 4}\,m_7^{-1}\,d\varth_{(1)}-{1\over 6}\,\xi_{(1)}\wedge\omg_{(1)}\,,\\[6 pt]
\varth^m S_m & \eql {1\over 12}\,m_7^{-1}\xi\,d\omg_{(1)}-{1\over 4}\,m_7^{-1}\,d\varth_{(1)}-{1\over 6}\,\xi_{(1)}\wedge\omg_{(1)}\,,
\end{split}
\end{equation}
we find that the internal potential is simply given by 
\begin{equation}\label{A3inv}
A_{(3)}^{\rm tr}\eql \alpha_1 \,S_{(3)}+\alpha_2\,\xi_{(1)}\wedge \omg_{(1)}\wedge \varth_{(1)}\,,
\end{equation}
where  
\begin{equation}\label{}
\begin{split}
 \alpha_1 & \eql  -{1\over 2\,\Sigma}  \sin  \zeta  \sinh (2 \lambda ) \,,\qquad 
 \alpha_2   \eql -{1\over 384\,\,\DX\,\Sigma}{\sin(2\zeta)\sinh^2(2\lambda)\over\sin^2\chi}\,.
\end{split}
\end{equation}
Rewriting \eqref{A3inv} in local coordinates using \eqref{oneforms} and \eqref{S3loc} yields
\begin{equation}\label{A3trloc}
A_{(3)}^{\rm tr}\eql {1\over 2}\,m_7^{-3}\,{\sin\zeta\sinh(2\lambda)\over\Sigma}\,\Big[J_{\CC\PP^3}-{1\over 2}\sin(2\chi)\,{\Xi\over \DX}\,d\chi\wedge d\phi\Big] \wedge \vartheta_{S^7}\,.
\end{equation}

Note that the  $A_{(3)}^{\rm tr}$  has only components along the internal manifold, $\cals M_7$, so that its field strength, $F_{(4)}^{\rm tr}$, can have at most one leg along the four-dimensional space-time. 

\subsection{The space-time flux}
\label{subSec:stflux}

We now turn to the second part of the flux, $F_{(4)}^{\rm st}$, which, as shown recently in  \cite{Godazgar:2013dma,Godazgar:2013pfa}, can be determined from the uplift for the transverse dual potential, $A_{(6)}^{\rm tr}$.  

The starting point is the Maxwell equation \eqref{maxwell} in eleven dimensions, which by setting $F_{(4)}=dA_{(3)}$ can be written locally as\footnote{See appendix~\ref{Appendix:A} for definitions and properties of the various duals used in this section.}
\begin{equation}\label{}
d(\star\,F_{(4)}+A_{(3)}\wedge F_{(4)})\eql 0\,,\end{equation}
from which  the dual potential, $A_{(6)}$,  is defined   by
\begin{equation}\label{}
dA_{(6)}\eql \star\,F_{(4)}+A_{(3)}\wedge F_{(4)}\,.
\end{equation}
The space time flux, $F_{(4)}^{\rm st}$,  is determined by the transverse part of $A_{(6)}$, that is \begin{equation}\label{Fst}
F_{(4)}^{\rm st}\eql -\star\,\big(dA_{(6)}^{\rm tr}-A_{(3)}^{\rm tr}\wedge F_{(4)}^{\rm tr}\big)\,,
\end{equation}
where 
\begin{equation}\label{}
A_{(6)}^{\rm tr}\eql {1\over 16}\,T_{(6)}-3 m_7\,\zetao_{(6)}\,.
\end{equation}
The six-form, $T_{(6)}={1\over 6!}T_{m_1\ldots m_6}dy^{m_1}\wedge\ldots \wedge dy^{m_6}$, is given by the uplift Ansatz  
\begin{equation}\label{Tuplift}
T_{m_1\ldots m_6}\eql \Delta g_{p[m_1}\,K_{m_2\ldots m_6]}{}^{IJ}K^{pKL}(u_{MN}{}^{IJ}+v_{MNIJ})(u^{MN}{}_{KL}+v^{MNKL})\,,
\end{equation}
where $K_{m_1\ldots m_5}^{IJ}\equiv i\,\bar\eta^I\Gamma_{m_1\ldots m_5}\eta^J$, while $\zetao_{(6)}$ is the potential for the volume of the round $S^7$, 
\begin{equation}\label{vols7}
d\zetao_{(6)}~\equiv~ {\rm v\oo l}_{S^7}\eql {1\over 8}\,m_7^{-7}\sin\chi\cos^5\chi\sin^3\theta\cos\theta\,
d\chi\wedge d\theta\wedge \sigma_1\wedge\sigma_2\wedge\sigma_3\wedge d\psi\wedge d\phi \,.
\end{equation}
Evaluating \eqref{Tuplift}, we find 
\begin{equation}\label{}
T_{(6)}\eql {8\over 3+\xi-6\coth(2\lambda)\sec\zeta}\,\sto_7\,\xi_{(1)}\,,
\end{equation}
where $\sto_7$ is the dual on $S^7$ with respect to the round metric. In terms of the local coordinates,
\begin{equation}\label{}
T_{(6)} \eql 2m_7^{-6}\,{\sin^2\chi\cos^6\chi\sin^3\theta\cos\theta\over  \cos(2\chi)-\coth(2\lambda)\sec\zeta}\,d\theta\wedge \sigma_1\wedge\sigma_2\wedge\sigma_3\wedge d\psi\wedge d\phi \,.
\end{equation}
Then
\begin{equation}\label{dT}
\begin{split}
dT_{(6)} & = -{4m_7^{-1}\, {\rm csch}^2(2\lambda) \sec^2\zeta \sin(2\chi)\over (\coth(2\lambda)\sec\zeta-\cos(2\chi))^2}
\big[4\cos\zeta\, d\lambda-\sin\zeta\sinh(4\lambda)\,d\zeta\big]\wedge \imath_{\partial_\chi}
{\rm v\oo l}_{S^7}\\[10 pt]
&\qquad + 8\,m_7\,\frac{  4 \coth (2 \lambda ) \sec  \zeta (1-2 \cos (2 \chi ))-4 \cos (2 \chi )+3 \cos (4 \chi
   )+5}{\left(  \coth (2 \lambda )\sec  \zeta -\cos (2 \chi )\right)^2}\,{\rm v\oo l}_{S^7}\,.
\end{split}
\end{equation}
From  \eqref{A3trloc}, we have 
\begin{equation}\label{A3F4}
A_{(3)}^{\rm tr}\wedge F_{(4)}^{\rm tr}\eql -m_7 \, \frac{\sin ^2\zeta\,\big[ (1-2 \cos (2 \chi ))\cos \zeta +3 \coth (2 \lambda )\big]}{(\cos
   \zeta +\coth (2 \lambda )) (\coth (2 \lambda )-\cos \zeta  \cos (2 \chi ))^2}\,
{\rm v\oo l}_{S^7}\,.
\end{equation}
Substituting \eqref{dT}, \eqref{vols7} and \eqref{A3F4} in \eqref{Fst}, we get
\begin{equation}\label{Fstst}
\begin{split}
F_{(4)}^{\rm st} &= \star\Big[{m_7^{-1}\,\sin(2\chi)\over (\cos\zeta\sinh(2\lambda)\cos(2\chi)-\cosh(2\lambda))^2}\,\big(\cos\zeta\, d\lambda-{1\over 4}\sin\zeta\sinh(4\lambda)\,d\zeta\big)\wedge \imath_{\partial_\chi} {\rm v\oo l}_{S^7}\\[10 pt]
&\hspace{50 pt} -{m_7\over \DX\Sigma^2}\big(\cos \zeta  \sinh (2 \lambda ) (2 \cos (2 \chi )-1)-3 \cosh (2 \lambda )\big)\, {\rm v\oo l}_{S^7}\Big]\,,
\end{split}
\end{equation}
where the dual is with respect to the full metric \eqref{d11met}. Using  identities
\eqref{decdual} and \eqref{stdual}  in appendix~\ref{Appendix:A} and 
\begin{equation}\label{}
\star\, {\rm v\oo l}_{S^7}\eql  \Delta^{-1}\,\star {\rm vol}_{M_7}\eql- \Delta^{-3}\,{\rm v\oo l}_{1,3}\,,\end{equation}
we find that the space time flux \eqref{Fstst} is
\begin{equation}\label{stFstar}
\begin{split}
F_{(4)}^{\rm st} & \eql -{m_7^{-1}\,\sin(2\chi)}\big(\cos\zeta\, \sto_{1,3}d\lambda-{1\over 4}\sin\zeta\sinh(4\lambda)\, \sto_{1,3}d\zeta\big)\wedge d\chi\\[6 pt]
& \qquad\qquad + {m_7\Delta^{-3}\over \DX\Sigma^2}\big(\cos \zeta  \sinh (2 \lambda ) (2 \cos (2 \chi )-1)-3 \cosh (2 \lambda )\big)\, {\rm v\oo l}_{1,3}\,.
\end{split}
\end{equation}

For the flow solutions where the scalar fields depend only on the radial coordinate, $r$,  we have $d\lambda=\lambda'\, dr$, $d\zeta=\zeta' \,dr$ and \eqref{stFstar} evaluates to a very simple expression,
\begin{equation}\label{spflux}
F_{(4)}^{\rm st} ~=~ {m_7\over 3 }\,e^{3A}\,{\rm v\oo l}_{1,2}\wedge  (U\,dr+V\,d\chi)\,,
\end{equation}
where
\begin{equation}\label{UandV}
\begin{split}
U(r,\chi) & ~=~ -3(1-2\cos 2\chi)\sinh 2\lambda \cos\zeta-9\cosh 2\lambda\,,\\[6 pt]
V(r,\chi) & ~=~  {3\over 4m_7^2}\,\sin 2\chi\,(4\cos\zeta\,\lambda'-\sinh (4\lambda)\sin\zeta\,\zeta')\,,
\end{split}
\end{equation}
and ${\rm v\oo l}_{1,2}$ is the volume along the $\textsl{Min}_{1,2}$ or $AdS_3$ slices.

It is straightforward to verify that $F_{(4)}^{\rm st}$ given in  \eqref{spflux} satisfies the Bianchi identity, $dF_{(4)}^{\rm st}=0$,  when the four-dimensional fields, $A(r)$, $\lambda(r)$ and $\zeta(r)$, are on-shell, that is they satisfy the equations of motion  \eqref{d4eqs} in four dimensions.

The calculation above illustrates  the point we have raised before, namely, that a rather long and complicated derivation using uplift formulae yields a relatively simple final result. In fact, after we have completed this calculation a paper \cite{Godazgar:2015qia} appeared where a more direct  Ansatz for the Freund-Rubin flux, namely the  term in $F_{(4)}^{\rm st}$ proportional to the volume of the four-dimensional space-time, is proposed. In the present context, the key observation is that the second term in  $U$ in \eqref{UandV} is the scalar potential \eqref{d4Pot} of the four-dimensional theory, while the first term is proportional to a derivative of the potential. This can be generalized to a more efficient uplift formula, which is summarized in 
appendix~\ref{Appendix:FR}.

\subsection{A summary of the uplift}
\label{subSec:theuplift}

We conclude this section with a brief summary of the eleven-dimensional fields constructed   in sections~\ref{subSec:metric}--\ref{subSec:stflux}. While the formulae for the uplifted fields are valid for any field configuration in four dimensions, here we will specialize them to the four-dimensional flows  we are interested in. 
 It turns out that the simplest form of the flux is obtained when we use suitable frames for the metric \eqref{d11met}. We will also need those frames later  in the proof of   supersymmetry of the RG flows and Janus solutions in section~\ref{Sec:susy11}. 

 Given   \eqref{d11met} and \eqref{theintmet}, a natural choice for the frames, $e^M$, $M=1,\ldots,11$, is to set
\begin{equation}\label{d11frames}
\begin{split}
e^{1,2,3} & \eql \DX^{1\over 6}\Sigma^{1\over 3}\,e^{A}\, f^{1,2,3}\,, \qquad 
e^4   \eql \DX^{1\over 6}\Sigma^{1\over 3}\,dr\,,\\[6 pt] 
e^5  & \eql m_7\,\DX ^{-\frac{1}{3}} \, \Sigma^{\frac{1}{3} } \, d\chi\,,\\[6 pt]
e^6 &  \eql m_7\, \DX ^\frac{1}{6} \, \Sigma^{-\frac{1}{6}} \, \cos\chi \, d\theta \,,\\[6 pt]  
e^{7,8}   & \eql   \frac{m_7}{2}\, \DX ^\frac{1}{6} \, \Sigma^{-\frac{1}{6}} \, \cos\chi \,\sin \theta \,  \sigma_{1,2}  \,,    \\[6 pt]
e^9 & \eql   \frac{m_7}{2}\, \DX ^\frac{1}{6} \, \Sigma^{-\frac{1}{6}} \,  \cos\chi \,\sin \theta \,\cos \theta \, \sigma_3 \,,   \\[6 pt]
e^{10} & \eql m_7 \, \DX ^\frac{2}{3} \, \Sigma^{-\frac{2}{3}} \,  \sin \chi \,  \cos\chi \,\Big((d\psi + \coeff{1}{2}\, \sin^2 \theta\, \sigma_3) +  \frac{\Xi}{\DX } \, d \phi  \Big) \,,  \\[6 pt]
e^{11} & \eql m_7 \, \DX ^{-\frac{1}{3}} \, \Sigma^{-\frac{2}{3}} \,  \big(d \phi + \cos^2 \chi \,(d\psi + \coeff{1}{2}\, \sin^2 \theta\, \sigma_3)  \big) \,,
\end{split}
\end{equation}
where $f^i$, $i=1,2,3$, are the frames for the $\textsl{Min}_{1,2}$ or $AdS_3$ slices, 
\begin{equation}\label{specfnct}
\begin{split}
\DX(r)\eql \cosh(2\lambda)& +\cos\zeta\,\sinh(2\lambda)\,,\qquad \Xi(r)\eql 2\cos\zeta\sinh(2\lambda)\,, \\[6 pt]
\Sigma(r,\chi) & \eql \cosh(2\lambda)-\,\cos\zeta\sinh(2\lambda)\cos(2\chi)\,.
\end{split}
\end{equation}
Then the transverse potential, $A_{(3)}^{\rm tr}$, given in \eqref{A3trloc} becomes surprisingly simple,
\begin{equation}\label{A3potfr}
A_{(3)}^{\rm tr}\eql {1\over 2} \,p(r)\,(e^6\wedge e^9+e^7\wedge e^8-e^5\wedge e^{10})\wedge e^{11}\,,
\end{equation}
where 
\begin{equation}\label{pofx}
p(r)\eql \sinh(2\lambda)\,\sin\zeta\,.
\end{equation}
Note that the coefficient function, $p(r)$, depends only on the four-dimensional space time radial coordinate. All dependence in \eqref{A3potfr} on the internal geometry and coordinates enters only through the frames.

Finally, the space time flux is given in \eqref{spflux} and \eqref{UandV}. This completes the constriction of the uplift.
 
\section{The equations of motion}
\label{Sect:EOMS}

In this section we verify explicitly that the metric  and the four-form flux in the uplift satisfy the equations of motion of eleven-dimensional supergravity when the four-dimensional metric and the scalar fields satisfy the four-dimensional equations of motion \eqref{d4eqs}--\eqref{4dEng}.

\subsection{Preliminaries}
\label{subSec:Prelim}

We start with some technical preliminaries that will help us  simplify the algebra in the calculations that follow. The main idea is to work directly with the functions that appear in the metric \eqref{theintmet}  and the flux \eqref{A3potfr}, in particular,  with  $\DX(r)$ and $p(r)$   given in \eqref{specfnct} and \eqref{pofx}, respectively, rather than with the scalar fields, $\lambda(r)$ and $\zeta(r)$.  To this end we use
\begin{equation}\label{elzeta}
\sin\zeta\eql -p\,\mathop{\rm csch}(2\lambda)\,,\qquad \cos\zeta\eql -\mathop{\rm csch}(2\lambda)(\cosh(2\lambda)-\DX)\,,
\end{equation}
and
\begin{equation}\label{ellambda}
\cosh(2\lambda)\eql {1+p^2+\DX^2\over 2\DX}\,,
\end{equation}
to eliminate $\zeta$ and $\lambda$ in terms of $p$ and $\DX$. This    converts complicated trigonometric expressions  into rational functions of the new fields $p$ and $\DX$ that are typically   easier to evaluate and simplify.  In particular,  the four-dimensional equations of motion \eqref{d4eqs}--\eqref{4dEng} in the rationalized form are given by
\begin{equation}\label{rsubddfs}
\begin{split}
\Delta'' & \eql -3\,A' \Delta' +\frac{1}{X}\Big[\left(p^2+1\right) \left(X'\right)^2+X^2 \left(p'\right)^2-2 p X p' X'\Big]
-2 \,g^2 \left(p^2+X^2-1\right)\\[10 pt]
p'' & \eql -3 \,p'A'+\frac{p}{X^2}{ \Big[\left(p^2+1\right) \left(X'\right)^2+X^2 \left(p'\right)^2-2 p X p'
   X'\,\Big]}-2\, g^2\frac{ p}{X} {\left(p^2+X^2+1\right)}\,,\\[10 pt]
A'' & \eql -{3\over 2}\,(A')^2-{3\over 8\DX^2}\,\Big[\left(p^2+1\right) \left(X'\right)^2+X^2 \left(p'\right)^2-2 p X p' X'\Big]\\[6 pt]
&\hspace{200pt} 
+{3\,g^2\over 2\DX}\,\left(1+(p')^2+(\DX')^2\right)-{e^{-2A}\over 2\ell^2} \,,  
\end{split}
\end{equation}
and
\begin{equation}\label{rateng}
(A')^2 -{1\over 4\DX^2}\,\Big[ \left(p^2+1\right) \left(X'\right)^2+X^2 \left(p'\right)^2-2 p X p' X'\Big]
-{g^2\over \DX}\left( p^2+\DX^2+1\right)
+{e^{-2 A}\over\ell^2}\eql 0\,.
\end{equation}
Similarly, we find that the superpotential 
\eqref{superW} is given by 
\begin{equation}\label{ratW}
W^2\eql {1\over 8X}\,\Big[\,{9\, p^4-6 \,p^2 \left(X^2-3\right)+\left(X^2+3\right)^2}\,\Big]
\end{equation}
and the BPS equations \eqref{JanusBPS:1} and \eqref{JanusBPS:2} for the scalars become
\begin{equation}\label{ratBPS}
\begin{split}
\DX' & \eql -\frac{1}{4\,W^2} \left[9 p^4-6 p^2 \left(X^2-1\right)+X^4+2 X^2-3\right]\,A'+\frac{e^{-A}}{\ell }\,
   {p \left(3 p^2-X^2+3\right)\over W^2}\,,\\[10 pt]
   p'& \eql -{ p\over 4\, \DX\,W^2}\, \left[9 p^4-6 p^2 \left(X^2-3\right)+X^4-2 X^2+9\right]\,
   A'\\[6 pt]
& \hspace{150 pt}+{e^{- A}\over\ell}\,{3 p^4+2 p^2 \left(X^2+3\right)-X^4-2 X^2+3\over 4\DX W^2}\,.
\end{split}
\end{equation}
As a consistency check one can verify once more that the first order equations \eqref{ratBPS} and 
 \eqref{Apsq} with $W$ given in \eqref{ratW} imply the second order equations \eqref{rsubddfs} and that \eqref{rateng} is equivalent to \eqref{Apsq}.
 
Finally, the other metric and the flux functions are:
\begin{equation}\label{SXrat}
\Sigma\eql {1\over \DX}\left[\,\cos^2\chi(1+p^2)+\sin^2\chi\,\DX\right]\,,\qquad 
\Xi\eql -{1\over\DX}\,\left(p^2-\DX^2+1\right)\,,
\end{equation}
and  
\begin{equation}\label{}
\begin{split}
U & \eql -{6\over \DX}\,\cos^2\chi \,(1+p^2)+3\DX(\cos(2\chi)-2)\,,\\[10 pt]
V & \eql  {3 \over 4 m_7^2}\,\sin 2\chi\,\Big[\,2(1+p^2)\,{\DX'\over\DX}-2p\,p'\Big]\,.
\end{split}
\end{equation}
This shows that indeed both the metric and the flux can almost entirely be written down using, up to overall factors, only rational functions of $\DX$ and $p$, and their derivatives!  

Finally, we will be often able to eliminate trigonometric functions of $\chi$ using
\begin{equation}\label{}
\cos(2\chi)\eql -\frac{p^2+X^2-2 \,\Sigma  X+1}{p^2-X^2+1}\,,
\end{equation}
which follows from \eqref{SXrat}.

\subsection{The flux}
\label{subSec:Flux}
 
The first place where using the rationalized parametrization becomes clearly advantageous is  the calculation of the components, $F_{MNPQ}$, of the four-form flux, $F_{(4)}$. Indeed, for the space-time part of the flux given in \eqref{spflux} we simply have
\begin{equation}\label{spflux}
\begin{split}
F_{1234}& \eql {m_7^{-1}\over\Sigma^{4/3}\DX^{2/3}}\,(2\,\Sigma+\DX)\,,\\[10 pt]
F_{1235}& \eql {\tan\chi\over \DX^{7/6}\Sigma^{4/3}\Xi}\,(\Sigma -X) \Big[\left(p^2+1\right) X'-p X p'\Big]\,.
\end{split}
\end{equation}

Turning to the transverse flux, $F_{(4)}^{\rm tr}=dA_{(3)}^{\rm tr}$, we note that
 the part of the three-form potential  along $\CC\PP^2$ in \eqref{A3trloc} has the complex structure, $J_{\CC\PP^2}$, as a factor. Thus the corresponding components of the field stength must satisfy
\begin{equation}\label{}
F_{MN69}\eql F_{MN78}\,.
\end{equation}
Modulo this identity, the non-vanishing components of the transverse part of the flux are:
\begin{equation}\label{intflux}
\begin{split}
F_{4510\,11} &\eql \frac{p X'-X p'}{2  X^{7/6}\Sigma^{1/3}}\,,\qquad 
F_{469\,11}  \eql \frac{\Sigma  p'-p \Sigma '}{2  \DX^{1/6}\Sigma ^{4/3}}\,,\\[10 pt]
F_{569\,10} & \eql -\frac{m_7 p (\Sigma +X)}{X^{2/3}\Sigma ^{4/3} }\,, \qquad 
F_{6789}  \eql \frac{2 m_7 p}{ X^{2/3}\Sigma^{1/3}}\,,
\end{split}
\end{equation}
where
\begin{equation}\label{dSdr}
\begin{split}
\Sigma'~\equiv~{\partial\Sigma\over\partial r} &\eql {1\over \DX^3-(p^2+1)\DX}\,\Big[
2 X \left(p X p'-(p^2+1)X' \right)\\[6 pt] &\hspace{150 pt}+\Sigma  \left(p^2 X'-2 p X
   p'+\left(X^2+1\right) X'\right)
\Big]\,.
\end{split}
\end{equation}
Later we will also need
\begin{equation}\label{dSdc}
{\partial\Sigma\over\partial \chi}\eql \sin(2\chi)\,\Xi\,.
\end{equation}
It appears that the flux produced through the uplift formulae is rather special, in particular, we 
find that the following components
\begin{equation}\label{}
F_{4569}\eql F_{469\,10}\eql F_{569\,11}\eql 0\,,
\end{equation}
accidentally vanish, that is not due to the underlying \suthuu\ symmetry of the construction.
   
\subsection{The Einstein equations}
\label{subSec:Eineqs}

The Einstein equations of eleven-dimensional supergravity in our conventions\footnote{See  appendix~\ref{Appendix:A}.} are:
\begin{equation}\label{Eineqs11}
R_{MN}+g_{MN}R  \eql {1\over 3}\,F_{MPQR}F_{N}{}^{PQR}\,.
\end{equation}
We start by evaluating the components of the Ricci tensor, $R_{MN}$, in the basis of   frames  \eqref{d11frames}. 
The  symmetries of the metric and the dependence of the scalar fields and the metric function in four dimensions     on the radial coordinate only, imply that   the non-vanisnhing  components of the Ricci tensor can be at most the following ones:
\begin{equation}\label{}
\begin{split}
R_{11}\eql -R_{22}\eql-R_{33}\,,\qquad & R_{44}\,,\qquad R_{45}\eql R_{54}\,,\qquad  R_{55}\,,\\
R_{66}\eql R_{77}&\eql R_{88}\eql R_{99}\,,\\
 R_{10\,10}\,,\qquad R_{10\,11}&\eql R_{11\,10}\,,\qquad R_{11\,11}\,.
\end{split}
\end{equation}
This agrees with the explicit result. Indeed, we find that
after imposing  the four-dimensional equations of motion \eqref{rsubddfs}--\eqref{rateng} in the
 rational parametrization introduced above, the diagonal components of the Ricci tensor  can be written in the form
 \begin{equation}\label{ricform}
R_{MM}\eql \cals A_{M}\,(\DX')^2+\cals B_{M}\,(p')^2+\cals C_{M}\,p'\,\DX'+\cals D_{M}\,,
\end{equation}
where $\cals A$, $\cals B$, $\cals C$ and $\cals D$ are functions of $p$, $X$ and $\chi$ (or, equivalently, $\Sigma$).  In particular, we find that the cross-terms
$A'\DX'$ and $A'p'$ are absent. Similarly, the off-diagonal components are of the form
\begin{equation}\label{ricoff}
R_{MN}\eql \cals A_{MN}\DX'+\cals B_{MN} \,p'+\cals D_{MN}\,,\qquad M\not=N\,.
\end{equation}
Explicit formulae for all non-vanishing coefficient functions are given in  appendix~\ref{Appendix:C}.

Evaluating the energy-momentum tensor on the right hand side in \eqref{Eineqs11} is straightforward. 
We will forego the details  and just look at one specific equation, 
the off-diagonal Einstein equation \eqref{Eineqs11} with $M=10$ and $N=11$. On the one side, we have\begin{equation}\label{}
R_{10\,11}\eql -2 \left(g^2-2 m_7^2\right) \tan  \chi \,\frac{ (\Sigma -X)}{X^{1/3}\Sigma ^{5/3} }\,.
\end{equation}
However, as one can see by inspection of the non-vanishing flux components,   the other side must be zero. This verifies the relation between the four-dimensional coupling constant, $g$, and the inverse radius of the internal manifold, $m_7$, \cite{de Wit:1986iy}
\begin{equation}\label{gandm}
g\eql\sqrt 2\,m_7\,.
\end{equation}
Given this relation, it is easy to check that all the remaining Einstein equations are satisfied as expected. 

\subsection{The Maxwell equations}
\label{subSec:Maxeqs}

The Maxwell equations  are
\begin{equation}\label{maxwell}
d\star F_{(4)}+F_{(4)}\wedge F_{(4)}\eql 0\,.
\end{equation}
For the flux \eqref{spflux}--\eqref{intflux},  they yield seven independent equations: four first order and three second order. 

The first order equations are along the components $[1234569\,11]$, $[1234578\,11]$, $[12356789]$, and  $[456789\,10\,11]$, and all have the same structure as this last one
\begin{equation}\label{}
\begin{split}
{4 m_7 p\tan\chi\over g^2\DX^{17/6}\Sigma^{5/3}\Xi^2}\, \left(2 m_7^2-g^2\right) \left(p^2-X^2+1\right) (\Sigma -X) \left(p^2 X'-p X
   p'+X'\right)\eql 0\,,\end{split}
\end{equation}
namely, they come with an overall factor of $(g^2-2m_7^2)$. 

The second order equations come from the components $[1234569\,10]$, $[1234578\,10]$ and $[12346789]$ in \eqref{maxwell}.
The first two equations are somewhat involved, but the last one is quite simple. There we find
\begin{equation}\label{}
\begin{split}
0 
& \eql {1\over 2\DX^{4/3}\Sigma^{2/3}}\,\Big[\,p\DX''-Xp''+3A'\,\left(pX'-Xp'\right)-24\,m_7^2
\,\Big] \\[10 pt]
& \eql \frac{2 p \left(g^2-2 m_7^2\right)}{ X^{4/3}\Sigma ^{2/3}}\,,
\end{split}
\end{equation}
where in going from the first to the second line we have used the four-dimensional equations of motion \eqref{rsubddfs}. Similarly, upon using \eqref{rsubddfs}, the other two equations reduce to the same expression modulo an overall factor of $\DX \Sigma$. Thus the Maxwell equations are satisfied if \eqref{gandm} holds.

To summarize, we have shown explicitly that the metric, $g_{MN}$, and the four-form flux, $F_{(4)}$, constructed using the uplift formulae in section~\ref{Sect:uplift} indeed satisfy the equations of motion of the eleven-dimensional supergravity when the scalar fields, $\lambda(r)$ and $\zeta(r)$, and the metric function,  $A(r)$, are on-shell in four dimensions. It is important  to note that  to verify that we have used only the  equations of motion in four dimensions \eqref{d4eqs}--\eqref{4dEng} or, equivalently, \eqref{rsubddfs}--\eqref{rateng}, but not the BPS equations! This means that also non-supersymmetric solutions of the same type will uplift to solutions of M theory. 
\section{Supersymmetry}
\label{Sec:susy11}

We now turn to  the  Janus and RG-flow solutions   of the BPS equations 
\eqref{JanusBPS:1}--\eqref{Apsq} and \eqref{RGA}--\eqref{RGflows}, respectively, to demonstrate explicitly the $\Neql (0,2)$ and  $\Neql 2$ supersymmetry of the corresponding uplifts in M theory. This has been discussed already in some detail in \cite{Pilch:2015vha}, where we have argued that the $\Neql 2$ supersymmetry of the RG flows is achieved by brane polarization and is naturally defined through projectors that reflect the underlying almost-complex structure and a dielectric projector much like those encountered in   \cite{Gowdigere:2003jf,Pilch:2004yg,Nemeschansky:2004yh,Pope:2003jp}. The  defect in the Janus solutions leads to additional chiral projector that is also present in four dimensions. The result for the RG-flows is then recovered by keeping both chiralities and taking the $\ell\to\infty$ limit. 

\subsection{Projector Ans\"atze}
\label{subSec:proj}

The BPS equations  in  eleven  dimensions are obtained by setting  the supersymmetry variations of the gravitinos to zero, 
\begin{equation}\label{d11BPS}
\delta\psi_M~\equiv~\partial_M\epsilon+\cM_M\,\epsilon\eql 0\,,
\end{equation}
where the algebraic operators, $\cM_M$, are given by 
\begin{equation}\label{}
\cals M_M~\equiv~  {1\over 4}\, \omega_{MPQ}\Gamma^{PQ}  +{1\over 144}\Big(\Gamma_M{}^{NPQR}-8\delta_M{}^N\Gamma^{PQR}\Big)\, F_{MNPQ}\,.
\end{equation}
The Killing spinors of unbroken supersymmetries are invariant under the Poincar\'e transformations in the  $tx$-plane and are singlets of    $\rm SU(3)$  acting along  $\CC\PP_2$. Hence $\epsilon$ does not depend on the cordinates $t$, $x$ and $\theta$,  as well as the Euler angles, $\alpha_1,\ldots,\alpha_3$. This means that the corresponding equations \eqref{d11BPS} are purely algebraic:\footnote{We will use the convention that the   indices $M=1,\ldots,11$ label components with respect to  the frames \eqref{d11frames}, while  $M=t,x,\ldots,\psi$, or $M=\sigma_1,\ldots,\sigma_3$, with respect to the local coordinates and/or forms.}
\begin{equation}\label{algeqs}
\cM_t\,\epsilon\eql\cM_x\,\epsilon\eql 0\,,\qquad \cM_\theta\,\epsilon \eql\cM_{\sigma_1}\eql\ldots\eql\cM_{\sigma_3}\eql 0\,.
\end{equation}
Similarly, the dependence of $\epsilon$ on the $\rm U(1)\times U(1)$ angles, $\phi$ and $\psi$, 
\begin{equation}\label{varphpsM}
{\partial\epsilon\over\partial\phi}\eql -\cM_\phi\,\epsilon\,,\qquad 
{\partial\epsilon\over\partial\psi}\eql -\cM_\psi\,\epsilon\,,
\end{equation}
is determined by the charges, $q_\phi=1 $ and $q_\psi=3/2$, respectively.   

Let us now consider the first equation in \eqref{algeqs}, written in the form
\begin{equation}\label{Meqs}
\frak M\,\epsilon\eql 0\,,\qquad \frak M~\equiv~\Gamma^1 \cM_1\,.
\end{equation}
The matrix $\frak M$, expanded into the basis of $\Gamma$-matrices, is given by
\begin{equation}\label{M1matga}
\begin{split}
\frak M & \eql {e^{-A}\over 2\ell}\,{1\over \DX ^{1/6}\Sigma^{1/3}}\,\Gamma^3 +{1\over 12\,\DX ^{7/6}\Sigma^{4/3}}\,\Big[2\DX {\partial\Sigma\over\partial r}+\Sigma\left(\DX '+6\DX  A'\right)\big]\,\Gamma^4 +{m_7 \DX ^{1/3}\over 6\Sigma^{4/3}}\,{\partial\Sigma\over\partial\chi}\,\Gamma^5\\[6 pt]
& \qquad +{1\over 3}\,\left(F_{1234}\Gamma^4+F_{1235}\Gamma^5\right)\,\Gamma^{123}+{1\over 6}\,F_{45\,10\,11}\,\Gamma^{45\,10\,11}+{1\over 6}\,F_{6789}\,\Gamma^{6789}\\[6 pt]
& \qquad +{1\over 6}\, \left(
F_{469\,11}\Gamma^{4\,11}+ F_{569\,10}\Gamma^{5\,10}\right)
\left(\Gamma^{69}+\Gamma^{78}\right)\,.
\end{split}
\end{equation}
Together with the explicit formulae \eqref{spflux}--\eqref{intflux}  for the flux, this gives us   a  homogenous system of linear equation for the thirty two components of $\epsilon$.

It is clear that after substituting the expressions for the flux components \eqref{spflux}--\eqref{intflux} and expanding the derivatives of $\Sigma$, see \eqref{dSdr} and \eqref{dSdc}, the operator $\frak M$, as well as the  other operators, $\cals M_M$, become  quite complicated. Hence, before we proceed  with the analytic calculation, we first explore numerically the space of solutions to \eqref{Meqs}. To do that, we first eliminate the derivatives $\DX'$, $p'$ and $A'$ using the BPS equations \eqref{ratBPS} and \eqref{Apsq}, and set $g=\sqrt 2\,m_7$. Next we assign random values to the   fields $\DX$, $p$, $A$, the angle $\chi$, and the constants $m_7$ and $\ell$ upon which  \eqref{Meqs} becomes a purely numerical system that can be solved for the components of  the Killing spinor, $\epsilon$. Note that our numerical assignment amounts simply to choosing random initial conditions for the four-dimensional BPS equations and thus is not constrained in any way. Those numerical solutions yield us some information about the subspace of allowed Killing spinors,  which confirms what one could also infer from an analysis in four dimensions and  the  \suthuu\ symmetry. More importantly, it allows us to short cut quite a bit of tedious analysis by fixing some of the signs in the projectors below that we would have to keep track of otherwise.

For finite $\ell$, the space of numerical solutions is generically two-dimensional in agreement with $\Neql (2,0)$ supersymmetry in four dimensions. The unbroken supersymmetries, $\epsilon$, must thus satisfy four  conditions 
\begin{equation}\label{projcond1}
\Pi_{0}\,\epsilon \eql \Pi_{1}\,\epsilon\eql\Pi_2\,\epsilon\eql \Pi_{3}\,\epsilon\eql 0\,,
\end{equation}
where $\Pi_0,\ldots,\Pi_3$ are mutually commuting projectors.
From the numerical analysis we  also find that two of these projectors are constant. To conform with the conventions in 
\cite{Pilch:2015vha}, we will denote them by $\Pi_1$ and $\Pi_3$. The first projector,
\begin{equation}\label{proj1}
\Pi_{1}  ~\equiv~{1\over 2}\,(1+\Gamma^{6789})\,,
\end{equation}
arises from the fact that the Killing spinor, $\epsilon$, must be a singlet under the holonomy group, $\rm SU(3)$, of $\CC\PP_2$. It depends on the choice of orientation of $\CC\PP_2$ defined by the frames $e^6,\ldots,e^9$. 
The second projector, 
\begin{equation}\label{proj3}
\Pi_{3}~\equiv~{1\over 2}\,(1-\Gamma^{12})\,,
\end{equation}
is just an uplift of the corresponding chirality projector  in four dimensions. In particular, choosing $\kappa=-1$ in  \eqref{JanusBPS:1} and    \eqref{JanusBPS:2} changes the sign in \eqref{proj3}.  Finally, we find that on the subspace of the Killing spinors satisfying   \eqref{projcond1},
\begin{equation}\label{Mphpseqs}
{\partial\epsilon\over \partial\phi}\eql -\Gamma^{69}\,\epsilon\,,\qquad 
{\partial\epsilon\over \partial\psi}\eql -{3\over 2}\,\Gamma^{69}\,\epsilon\,.
\end{equation}
Together with \eqref{varphpsM}, this gives us two additional algebraic equations, which as we will see simplifies the calculations significantly.
We should also note that both projectors do not depend on the choice of the square root branch in 
\eqref{Apsq} used to eliminate $A'$.  

For the RG flows, taking the limit $\ell\to\infty$ eliminates the first term in \eqref{M1matga}. The space of solutions includes then both $\Gamma^{12}$-chiralities;  the projector $\Pi_{3}$ is thus absent and we have a four-dimensional space of solutions corresponding to $\Neql 2$ supersymmetry. We have shown in \cite{Pilch:2015vha}  that   the remaining two commuting projectors in this limit are
\begin{equation}\label{proj2inf}
\Pi_2^{\infty}\eql {1\over 2}\,\Big[  \, 1+(\cos\alpha\,\Gamma^5-\sin\alpha\,\Gamma^4)\,\Gamma^{69}(\cos\omega \,\Gamma^{10}+\sin\omega\,\Gamma^{11})\, \Big]\,,
\end{equation}
and
\begin{equation}\label{proj0inf}
\Pi_0^{\infty}\eql {1\over 2}\,\Big[1+\cos\beta\,\Gamma^{123}+\sin\beta\,(\cos\alpha\,\Gamma^4+\sin\alpha\,\Gamma^5)\Big]\,,
\end{equation}
where the angles $\alpha$, $\beta$ and $\omega$ are  some functions of $r$ and $\chi$.\footnote{See,   (4.8) and (4.5)  in \cite{Pilch:2015vha}.} In analogy with \eqref{proj1}, the  projector \eqref{proj2inf} can be associated with an extension of the complex structure of $\CC\PP_2$ to an almost complex structure with  extra pair of complex frames. Finally, \eqref{proj0inf} is the dielectric deformation of the standard M2-brane projector at $\beta=0$.

For the  Janus solutions, the projectors \eqref{proj2inf} and \eqref{proj0inf}
must be deformed to account for the defect, which gives rise to additional terms in the supersymmetry variations, such as the first term  in~\eqref{M1matga}. Including such terms in 
\eqref{proj2inf} and  \eqref{proj0inf} leads to the following Ansatz for the projectors at finite $\ell$:
\begin{equation}\label{smproj}
\Pi_0\eql {1\over 2}\,\Big[ \, 1+a_1\Gamma^3+a_2\Gamma^4+a_3\Gamma^5\, \Big]\,,
\end{equation}
and
\begin{equation}\label{bigproj}
\Pi_2\eql {1\over 2}\Big[ \, 1+(b_1\Gamma^3+b_2\Gamma^4+b_3\Gamma^5)\,\Gamma^{69}(\cos\omega \,\Gamma^{10}+\sin\omega\,\Gamma^{11})\, \Big]\,.
\end{equation}
Those two operators  form a pair of commuting projectors provided the vectors 
 $\bfs a\equiv(a_1,a_2,a_3)$ and $\bfs b\equiv (b_1,b_2,b_3)$ are orthonormal. Such a pair of vectors can be parametrized by three angles, $\alpha$, $\beta$ and $\gamma$:
\begin{equation}\label{angas}
\begin{split}
a_1\eql \cos \beta  \cos \gamma  & -\sin \alpha   \sin \beta  \sin \gamma  \,,\qquad 
a_2\eql \cos \alpha \sin \beta \,,\\[6 pt]
a_3 & \eql \sin \alpha \sin \beta  \cos \gamma  +\cos \beta  \sin \gamma  \,,
\end{split}\end{equation}
and
\begin{equation}\label{angbs}
b_1\eql -\cos\alpha\sin\gamma\,,\qquad b_2\eql -\sin\alpha\,,\qquad b_3\eql \cos\alpha\cos\gamma\,.
\end{equation}
Together with $\omega$ those angles are some functions of $r$ and $\chi$ and will be determined by solving the supersymmetry variations. For $\gamma=0$, the projectors $\Pi_0$ and $\Pi_2$ reduce to 
$\Pi_0^\infty$ and $\Pi_2^\infty$, respectively. We can thus view the angle $\gamma$ as the Janus deformation parameter which goes to zero in the RG-flow limit.

There is still certain redundancy in our description of the projectors \eqref{smproj} and \eqref{bigproj}.  To see this, introduce a third vector,  $\bfs c$, so that $({\bfs a}, {\bfs b}, {\bfs c})$ are orthonormal and define ${\bfs x}\cdot \Gamma \equiv (x_1\Gamma^3+x_2\Gamma^4+x_3\Gamma^5)$.   Observe that the product  $({\bfs b}\cdot \Gamma)({\bfs c}\cdot \Gamma)\Gamma^{10}\Gamma^{11}$ commutes with all the projectors, $\Pi_0, \dots, \Pi_3$ and so preserves the space of supersymmetries. One is therefore free to rotate \eqref{smproj} and \eqref{bigproj} using the action of $({\bfs b}\cdot \Gamma)({\bfs c}\cdot \Gamma)\Gamma^{10}\Gamma^{11}$ and this induces a simultaneous rotation $\omega\to\omega+\vartheta$ accompanied by a rotation of  $\bfs b$ and $\bfs c$ by the angle $\vartheta$.  In the following we will use this freedom to simplify our calculations.

\subsection{Supersymmetries for the Janus solutions}
\label{subSec:solsuper}

We will now calculate all the projectors and the Killing spinor, $\epsilon$, by solving explicitly the BPS equations \eqref{d11BPS}. 

In principle, one should be able to determine all the projectors in \eqref{projcond1}, or equivalently  solve for the angles $\alpha,\,\beta\,,\gamma$ and $\omega$,  directly from \eqref{Meqs}. The problem is that this effectively amounts to obtaining the individual projectors $\Pi_0,\ldots,\Pi_3$ from a particular linear combination of products of these projectors.  This, unsurprisingly, is  not the best way to proceed.  Instead, we  will first solve algebraic equations that arise from   judicious linear combinations of the variations \eqref{d11BPS} in which   the flux terms  either cancel completely or are simple. 

The first such equation arises from the  ``magical combination'' of  variations 
\begin{equation}\label{}
2\,\Gamma^1\delta\psi_1+\Gamma^6\delta\psi_6+\Gamma^7\delta\psi_7+\Gamma^{10}\delta\psi_{10}+\Gamma^{11}\delta\psi_{11}\eql 0\,,
\end{equation}
in which all flux terms cancel. After eliminating the derivatives with respect to the $\rm U(1)$ angles using \eqref{Mphpseqs} and modulo terms annihilated by $\Pi_1$ and $\Pi_3$, it reads
\begin{equation}\label{dvar}
\big[\,\fA_1\,\Gamma^3 +\fA_2\,\Gamma^4+\fA_3\,\Gamma^5+ \Gamma^{69}\,(\fA_4\,\Gamma^{10}+\fA_5\,\Gamma^{11})\,\big]\,\epsilon\eql 0\,,
\end{equation}
where
\begin{equation}\label{defofA}
\begin{split}
\fA_1 & \eql {1\over \Sigma^{1/3}\DX^{1/6}}\,{e^{-A}\over \ell }\,,\qquad \fA_2\eql 
{A'\over \Sigma^{1/3}\DX^{1/6}}\,,\qquad 
\fA_3\eql \frac{ m_7\,\DX^{1/3} (2 \cos (2 \chi )-1) }{\sin(2\chi)\,\Sigma^{1/3}}\,,\\[6 pt]
\fA_4 & \eql -\frac{m_7\,{\DX^{1/3}} (\cos (2 \chi )-2)  )}{ \sin(2\chi)\, {\Sigma ^{1/3}}}\,,\qquad
\fA_5\eql \frac{m_7\,{\DX^{1/3}} \left(\DX (\cos (2 \chi )-2)+3 \,\Sigma \right)}{2   \cos^2\chi\,
    {\Sigma^{1/3} }}\,.
\end{split}
\end{equation}
Iterating \eqref{dvar} one finds a single  consistency condition
\begin{equation}\label{constcA}
\fA _1^2+\fA _2^2+\fA _3^2-\fA _4^2-\fA _5^2\eql0\,,
\end{equation}
which is satisfied by virtue of   \eqref{Apsq} and \eqref{gandm}. This condition also means that, up to an invertible factor, \eqref{dvar} is in fact the  projector \eqref{bigproj} with
\begin{equation}\label{bpars}
b_1\eql {\fA _1\over \fA }\,,\qquad b_2\eql {\fA _2\over \fA }\,,\qquad b_3\eql {\fA _3\over \fA }\,,\qquad \cos\omega\eql {\fA _4\over \fA }\,,\qquad \sin\omega \eql {\fA _5\over \fA }\,,
\end{equation}
where
\begin{equation}\label{}
\fA \equiv (\fA _1^2+\fA _2^2+\fA _3^2)^{1/2}\eql (\fA _4^2+\fA _5^2)^{1/2}\,.
\end{equation}
Using  \eqref{angbs} and \eqref{bpars}, we  then read off
\begin{equation}\label{alphaang}
\begin{split}
\cos\alpha\cos\gamma& \eql -{(2\cos(2\chi)-1) {\DX^{1/2}}\over\Omega^{1/2}}
\,,\qquad
\cos\alpha\sin\gamma\eql -{a\sin(2\chi)\over \Omega^{1/2}}\,{e^{-A}\over\ell}\,,
\\[6 pt]
& \hspace{1 in}\sin\alpha  \eql -{a\sin(2\chi)\,A'\over\Omega^{1/2}}\,,
\end{split}
\end{equation}
and
\begin{equation}\label{omegares1}
\cos\omega\eql  {(\cos(2\chi)-2)\DX^{1/2}\over\Omega^{1/2}}\,,\qquad \sin\omega\eql 
 {\sin(2\chi)(3\, p^2-\DX^2+3)\over 2\,\DX^{1/2}\Omega^{1/2}} 
\,,
\end{equation}
where
\begin{equation}\label{defOm}
\begin{split}
\Omega & \eql (1-2\cos(2\chi))^2\DX + 2\,\sin^2(2\chi)\,W^2\,.
\end{split}
\end{equation}

 Before proceeding we note that the rotation of the gamma matrices that define the projectors is equivalent to a rotation of the frames. In particular, the rotation by $\omega$ is equivalent to starting with the frames:   
\begin{equation}
\hat e^{10} ~\equiv~\cos\omega\, e^{10}+\sin\omega\, e^{11}\,, \qquad \hat e^{11} ~\equiv~-\sin\omega\, e^{10}+\cos\omega\, e^{11}\,.
\end{equation}
Using (\ref{omegares1}) we find a rather simple result for one of these frames:
\begin{equation}
\hat e^{10} ~=~   m_7 \, \DX^{\frac{1}{6}} \, \Sigma^{\frac{1}{3}} \, \Omega^{-\frac{1}{2}}\,    \big(d\phi ~+~ \coeff{3}{2}\,(d\psi + \coeff{1}{2}\, \sin^2 \theta\, \sigma_3)  \big)   \,.
\label{e10hatsimp}
\end{equation}
Note that the mixing of $\phi$ and $\psi$ does not involve functions of $r$ and furthermore (\ref{Mphpseqs}) implies that  the supersymmetries only depend upon angles in precisely the combination $(\phi + \frac{3}{2} \psi)$.  We will return to this observation later.

Continuing with the supersymmetry analysis, 
 since the projectors \eqref{smproj} and \eqref{bigproj} commute, we  cannot obtain any information from \eqref{dvar} about the dielectric polarization angle, $\beta$. For that we turn to another magical combination,
\begin{equation}\label{}
\Gamma^1\,\delta\psi_1+\Gamma^7\,\delta\psi_7+\Gamma^8\,\delta\psi_8\eql 0\,,
\end{equation}
which has no derivatives of $\epsilon$ and no terms with components of the internal flux. After imposing the constant projections, it reads
\begin{equation}\label{Bvar}
\big[\,\fB_1+\fB_2\,\Gamma^3 +\fB_3\,\Gamma^4+\fB_4\,\Gamma^5+ \Gamma^{69}\,(\fB_5\Gamma^{10}+\fB_6\Gamma^{11})\,\big]\,\epsilon\eql 0\,,
\end{equation}
where
\begin{equation}\label{Bdefs}
\begin{split}
\fB_1 & \eql \frac{m_7\,p }{\Sigma^{3/2} \DX^{2/3}}\,,\qquad 
\fB_2\eql {1\over 2\,\Sigma^{1/3}\DX^{1/6}}\Big[{e^{-A}\over \ell}-{p \DX'-p '\DX\over 2\DX}\Big]\,,\\[6 pt]
\fB_3   & \eql \frac{2 \DX A'+\DX'}{4\, \Sigma^{3/2} \DX^{7/6}}\,,  \qquad  \fB_4    \eql -\fB_5\eql -\frac{m_7\,{\DX^{1/3}} \tan \chi }{  \Sigma^{1/3}}\,,\qquad \fB_6\eql{m_7  \over \Sigma^{1/3}\DX^{2/3}}\,.
\end{split}
\end{equation}
Note that the presence of the $\fB_1$-term in \eqref{Bvar}, with an analogous term absent in \eqref{dvar}, prevents \eqref{Bvar} from being a projector. Still, by iteration one finds a consistency condition 
\begin{equation}\label{}
\fB_1^2-\fB_2^2-\fB_3^2-\fB_4^2+\fB_5^2+\fB_6^2\eql 0\,,
\end{equation}
which is indeed satisfied by virtue of \eqref{ratBPS} and \eqref{Apsq}.

Using the projectors \eqref{smproj} and \eqref{bigproj} in \eqref{Bvar}, one is left with three independent products of $\Gamma$-matrices which yield the following  equations:
\begin{equation}\label{beq1}
\begin{split}
  \fB_1 (\sin \alpha  \sin \beta  \cos \gamma +\cos \beta  \sin\gamma )+\left(\fB_5 \cos\omega +\fB_6 \sin \omega  \right) \cos \alpha  \cos \gamma & \\ 
+ \left(\fB_5 \sin \omega -\fB_6 \cos\omega \right) (\sin \alpha  \cos \beta  \cos \gamma -\sin \beta  \sin \gamma )-\fB_4 & \eql 0 \,,
\end{split}
\end{equation}
\begin{equation}\label{beq2}
\begin{split}
& \fB_2 (\sin \alpha \sin \beta  \cos \gamma  +\cos \beta  \sin  \gamma
    )+\fB_4 (\sin \alpha \sin \beta  \sin \gamma  -\cos \beta  \cos  \gamma
    )\\  &\qquad  +(\fB_5 \cos \omega +\fB_6 \sin \omega ) \cos \alpha \cos \beta   +(\fB_5 \sin \omega-\fB_6\cos \omega)  \sin \alpha   \eql 0\,,
\end{split}
\end{equation}
\begin{equation}\label{beq3}
\begin{split}
& \fB_3 (\sin \alpha        \sin \beta         \cos
   \gamma     +\cos \beta         \sin \gamma )    -\fB_4 \cos \alpha        \sin \beta    -  \left(\fB_5 \sin \omega       -\fB_6 \cos \omega       \right)   \cos \alpha        \sin \gamma 
  \\& \qquad  +
\left( \fB_5 \cos \omega + \fB_6 \sin \omega            \right) (\sin \alpha        
\cos \beta         \sin \gamma         +\sin \beta         \cos \gamma         )
   \eql 0\,. \end{split}
\end{equation}
However, only one of those equations is independent, which can be seen by solving one of them for  $\tan(\beta/2)$ and then verifying that the other two are satisfied. Equivalently, one can solve the first two for $\cos\beta$ and $\sin\beta$ and then check that their squares indeed add up to one. Substituting the result into the third equations yields a consistency condition  
\begin{equation}\label{}
\begin{split}
\fB_2 (\cos  \beta   \cos
    \gamma - \sin  \alpha  \sin \beta \sin \gamma)& +\fB_3 \cos  \alpha   \sin  \beta \\&   +\fB_4( \sin \alpha  \sin \beta
    \cos \gamma +  \cos  \beta   \sin  \gamma )   -\fB_1\eql 0\,.
\end{split}
\end{equation}
This equation has a simple  geometrical interpretation, namely that the (non-unit) vector  
\begin{equation}\label{}
\bfs d~\equiv~ \Big({\fB_2\over\fB_1},{\fB_3\over\fB_1},{\fB_4\over\fB_1}\Big)\,,
\end{equation}
satisfies
\begin{equation}\label{}
\bfs a\cdot \bfs d\eql -1\,,
\end{equation}
which is the consistency condition between the operator in \eqref{Bvar} and the projector \eqref{smproj}.

Similarly as for the equations of motion in section~\ref{Sect:EOMS}, 
  the solution for $\cos\beta$ and $\sin\beta$ above can be simplified using rationalized BPS equations. After some algebra, we  find the following result:
\begin{equation}\label{}
\cos\beta\eql \cos\gamma\,C_0+\sin\gamma\,C_1\,,\qquad \sin\beta\eql \cos\gamma\,S_0+\sin\gamma\,S_1\,,
\end{equation}
where 
\begin{equation}\label{}
\begin{split}
C_0 & \eql -{m_7^{-1}\over 4\,\DX  ^{3/2} \, \Sigma  }   
   \Big[   \left(\DX  ^2+3\right) \left(\DX  ^2 \sin ^2\chi +\cos ^2\chi \right)\\ &
  \hspace{100pt} +p^2
   \left(\DX  ^2 (\cos (2 \chi )-2)+6 \cos ^2\chi \right)+3 \,p ^4 \cos ^2\chi\,  \Big]\,{A'\over W^2}\,,\\[10 pt]
S_0 & \eql {p  \,\Omega^{1/2}\over \sqrt 2\,\,\Sigma}{A'\over W^2}\,,
\end{split}
\end{equation}
 and 
\begin{equation}\label{}
\begin{split}
C_1 & \eql \frac{p  \sin (2 \chi )}{8 \, \Sigma\, \DX  \, W^2}
 \Big[-\DX  ^4+ (16 \cos (2 \chi )-5 (\cos (4 \chi )+3)) \csc
   ^2(2 \chi )\DX  ^2  \\ & \hspace{250 pt}+ 6  p ^2\left(\DX  ^2-3\right)-9 p ^4-9\Big]\,, \\[10 pt]
S_1 & \eql -\frac{\sqrt{\Omega } \csc (2 \chi ) }{4 \,\Sigma \,\DX  ^{3/2} \,
   W^2}
\Big[\DX  ^4 \sin ^2(\chi )+\DX  ^2 \left(p ^2-1\right)
   (\cos (2 \chi )-2)+3 \left(p ^2+1\right){}^2 \cos ^2(\chi )\Big]\,.
   \end{split}
\end{equation}  
This completes the calculation of all the angles in the  projectors $\Pi_0$ and $\Pi_2$.

To determine explicitly the  Killing spinors for unbroken supersymmetries, let us introduce rotations 
\begin{equation}\label{}
\cals R_{ij}(x)\eql \cos x-\sin x\,\Gamma^{ij}\,,\qquad i,j>1\,,
\end{equation}
and define 
\begin{equation}\label{rotation}
\cals R(\alpha,\beta,\gamma,\omega)\eql \cals R_{35}(\gamma/2)\,\cals R_{45}(\alpha/2)\,\cals R_{34}(\beta/2)\,\cals R_{10\,11}(\omega/2)\,,
\end{equation}
which commute with the projectors $\Pi_1$ and $\Pi_3$. It is   straightforward to check that 
the projectors \eqref{smproj} and \eqref{bigproj} are then simply
\begin{equation}\label{}
\begin{split}
\Pi_0 & \eql \cals R(\alpha,\beta,\gamma,\omega)\,\Pi_0^{(0)}\,\cals R(\alpha,\beta,\gamma,\omega)^{-1}\,,\qquad 
\Pi_2 \eql \cals R(\alpha,\beta,\gamma,\omega)\,\Pi_2^{(0)}\,\cals R(\alpha,\beta,\gamma,\omega)^{-1}\,,
\end{split}
\end{equation}
where
\begin{equation}\label{zproj2}
\Pi_3^{(0)}\eql {1\over 2}(1+\Gamma^{3})\,,\qquad \Pi_4^{(0)}\eql {1\over 2}(1+\Gamma^{569\,10})\,.
\end{equation}
Thus any solution $\epsilon$ to \eqref{projcond1} can be written as
\begin{equation}\label{}
\epsilon\eql \cals R(\alpha,\beta,\gamma,\omega)\,\tilde \epsilon\,,
\end{equation}
where $\tilde \epsilon$ is in the kernel of the constant projectors \eqref{proj1}, \eqref{proj3} and \eqref{zproj2}. 

From the supersymmetry variations along the radial direction, $y$,  we find  
\begin{equation}\label{}
{\partial\tilde \epsilon\over \partial y}\eql {1\over 2\ell}\,\tilde \epsilon\,,
\end{equation}
which is the correct radial dependence for the Killing spinor along $AdS_3$. 

This leaves two variations along $r$ and $\chi$, which are solved as usual by setting
\begin{equation}\label{}
\tilde \epsilon\eql H_0^{1/2}\,\varepsilon\,,
\end{equation}
 where 
 \begin{equation}\label{}
H_0\eql \DX^{1/ 6}\Sigma^{1/3}\,e^{A(r)}\,,
\end{equation}
is the warp factor of the ``time'' frame, $e^1=H_0\,dt$, and $\varepsilon$ is a constant spinor along the internal manifold and with the standard dependence along $AdS_3$, which satisfies the same constant projections as $\tilde\epsilon$.
 
\subsection{The RG-flow limit}
\label{subSec:solsuper}

The supersymmetry analysis simplifies significantly for the holographic flow solution.  For this one simply imposes the projectors (\ref{proj1}),  (\ref{proj2inf}) and (\ref{proj0inf}) but does not impose a helicity projector like (\ref{smproj}).  We then find, taking the upper signs in (\ref{RGA}) and (\ref{RGflows}):
\begin{equation}\label{alphaflow}
\cos\alpha   \eql  {(2\cos(2\chi)-1) {\DX^{1/2}}\over\Omega^{1/2}} \,,\qquad  \sin\alpha  \eql -{a\sin(2\chi)\,A'\over\Omega^{1/2}}\,,
\end{equation}
\begin{equation}\label{omegaflow}
\cos\omega\eql  {(\cos(2\chi)-2)\DX^{1/2}\over\Omega^{1/2}}\,,\qquad \sin\omega\eql 
 {\sin(2\chi)(3\, p^2-\DX^2+3)\over 2\,\DX^{1/2}\Omega^{1/2}}  \,,
\end{equation}
and
\begin{equation}\label{}
\begin{split}
\cos\beta & \eql -
\frac{1}{2 \sqrt{2}\, W  \,\DX  ^{3/2} \, \Sigma  }   \,
   \Big[   \left(\DX  ^2+3\right) \left(\DX  ^2 \sin ^2\chi +\cos ^2\chi \right)\\ &
  \hspace{100pt} +p^2
   \left(\DX  ^2 (\cos (2 \chi )-2)+6 \cos ^2\chi \right)+3 \,p ^4 \cos ^2\chi\,  \Big]  \,,\\[10 pt]
\sin\beta  & \eql {p  \,\Omega^{1/2}\over \sqrt 2\,W\,\Sigma}\,.
\end{split}
\end{equation}
The space-time components of the Maxwell fields also simplify and we obtain a seemingly standard relation for holographic flows:
\begin{equation}\label{h0beta}
h_0  \eql -\frac{1}{2} \cos\beta \,.
\end{equation}
%

\section{IR asymptotics in eleven dimensions}
\label{Sec:LargeLambda}

Having constructed the uplift in detail, we now examine the infra-red limits of the holographic RG flows  described by  (\ref{RGA}) and (\ref{RGflows})  from the perspective of M theory.  In an earlier paper \cite{Pilch:2015vha}  we  focussed  upon the special flow with $\zeta = \pi/3 $ since this led to a very interesting new result.  Here we will complete the asymptotic analysis for all flows.

First recall that $\zeta$ limits to a constant value as $\lambda \to +\infty$ and so the various warp factors behave as follows:
\begin{equation}
 \DX  ~\sim~ \coeff{1}{2} \,(1 + \cos \zeta) \, e^{2 \lambda}\,, \qquad  \Xi  ~\sim~ \cos\zeta  \, e^{2 \lambda} \,,
 \qquad \Sigma  ~\sim~ \coeff{1}{2} \,  e^{2 \lambda} \, \widehat \Sigma \,, \qquad \widehat \Sigma \ ~\equiv~ (1 - \cos \zeta \, \cos 2\chi)
\label{11dasymp1}\,.
\end{equation}
%

\subsection{$\cos 3\zeta \ne  -1$}
\label{genzeta}

For such a generic $\zeta$ one has 
\begin{equation}
d \lambda  ~\sim~ \mp \frac{g}{4} \, \sqrt{(1 +\cos 3\zeta)} \, e^{3 \lambda}\, dr  \,, 
 \qquad e^A  ~\sim~ R^2 \,  e^{- \lambda}   \label{11dasymp2}\,,
\end{equation}
for some constant, $R$.   Thus the warp factor for the branes and the corresponding frames are finite and smooth for $\zeta \ne 0, \pi$:

\begin{equation}
e^i    ~\sim~ \frac{1}{ \sqrt{2}} \, R^2 \, (1 +\cos \zeta)^\frac{1}{6}  \,\widehat \Sigma^\frac{1}{3}   \, dx^i \,, \qquad i = 1,2,3  \,;  \label{11dasymp3}
\end{equation}
Thus the metric parallel to the branes is simply:
\begin{equation}
ds_3^2 ~=~\sum_{i=1}^{3} \ (e^i)^2 ~\sim~  (1 +\cos \zeta)^\frac{1}{3}  \,R^4\, \widehat \Sigma^\frac{2}{3}  \, (-dx_1^2 +dx_3^2 + dx_3^2)\  \,.\label{3met1}
\end{equation}

From   (\ref{d11frames})  and (\ref{11dasymp1}) one has 
\begin{equation}\label{11dasymp4}
\begin{split}
e^4   ~\sim~& \mp \frac{2 \sqrt{2}}{g}  \, \frac{(1 +\cos \zeta)^\frac{1}{6}\,\widehat \Sigma^\frac{1}{3} }{ (1 +\cos 3\zeta)^{\frac{1}{2}}} \,  e^{-2 \lambda} \, d \lambda  \,,  \\[6 pt]
 e^{11} ~\sim~&   \frac{2\, m_7}{(1 + \cos \zeta)^{\frac{1}{3}}  \,  \widehat \Sigma^{\frac{2}{3}}  }\, e^{-2 \lambda} \,  \big(d \phi + \cos^2 \chi \,(d\psi + \coeff{1}{2}\, \sin^2 \theta\, \sigma_3) \big) \,, 
\end{split}
\end{equation}
These are the only two frames to depend on $\lambda$ in this limit.

The remaining frames limit to:
\begin{align}
e^5  ~\sim~ & m_7\, \Big(\frac{\widehat \Sigma}{1 + \cos \zeta}\Big)^{\frac{1}{3}} \, d\chi \,; \qquad
e^6 ~\sim~   m_7\,\Big(\frac{\widehat \Sigma}{1 + \cos \zeta}\Big)^{-\frac{1}{6}} \, \cos\chi \, d\theta \,;  \label{genlimfram1} \\  
e^7 ~\sim~ &   \frac{m_7}{2}\, \Big(\frac{\widehat \Sigma}{1 + \cos \zeta}\Big)^{-\frac{1}{6}}\, \cos\chi \,\sin \theta \,  \sigma_1 \,;     \\
e^8  ~\sim~ &   \frac{m_7}{2}\, \Big(\frac{\widehat \Sigma}{1 + \cos \zeta}\Big)^{-\frac{1}{6}}\, \cos\chi \, \sin \theta \, \sigma_2 \,;  \label{genlimfram2} \\
 e^9 ~\sim~ & \frac{m_7}{2} \, \Big(\frac{\widehat \Sigma}{1 + \cos \zeta}\Big)^{-\frac{1}{6}} \,  \cos\chi \,\sin \theta \,\cos \theta \, \sigma_3 \,;   \label{genlimfram3} \\
\qquad e^{10} ~\sim~ & m_7 \,\Big(\frac{\widehat \Sigma}{1 + \cos \zeta}\Big)^{-\frac{2}{3}}\,  \sin \chi \,  \cos\chi \,\Big((d\psi + \coeff{1}{2}\, \sin^2 \theta\, \sigma_3) +  \frac{2 \, \cos \zeta }{(1+\cos \zeta )} \, d \phi  \Big) \,.  \label{genlimfram4}  
\end{align}
It is instructive to rewrite $e^{11}$ in terms of the one-form appearing in $e^{10}$:
\begin{equation}
 e^{11} ~\sim~   \frac{2\, m_7 \,  \widehat \Sigma^{\frac{1}{3}} }{(1 + \cos \zeta)^{\frac{4}{3}}   }\, e^{-2 \lambda} \,  \Big[d \phi+ \frac{(1+\cos \zeta ) \cos^2 \chi }{\widehat \Sigma }\,\Big(d\psi + \coeff{1}{2}\, \sin^2 \theta\, \sigma_3  +  \frac{2 \, \cos \zeta }{(1+\cos \zeta )} \, d \phi\Big) \Big] \,, \label{newe11asymp}
\end{equation}
Then one has
\begin{equation}
\begin{split}
ds_2^2 &  ~=~   (e^4)^2 ~+~ (e^{11})^2 \\ & ~\sim~   
 \frac{m_7^2 \, (1 +\cos \zeta)^\frac{1}{3}\,\widehat \Sigma^\frac{2}{3} }{ (1 +\cos 3\zeta) }\, \Big[d \rho^2   
  +  \rho^2\,  \frac{4\,(1 +\cos 3\zeta) }{(1 + \cos \zeta)^3  }\times \\[6 pt]
  & \hspace{70 pt} \times\, \big(d \phi+ \frac{(1+\cos \zeta ) \cos^2 \chi }{\widehat \Sigma }\,\Big(d\psi + \coeff{1}{2}\, \sin^2 \theta\, \sigma_3  +  \frac{2 \, \cos \zeta }{(1+\cos \zeta )} \, d \phi\big) \Big)^2 \Big] \,, \label{r2met}
\end{split}
\end{equation}
where $\rho \equiv e^{-2 \lambda}$.

The remaining part of the metric is
\begin{equation} \label{6met1}
\begin{split}
ds_6^2 ~=~ & \sum_{j=5}^{10} \ (e^j)^2 \\ ~\sim~   
& m_7^2\, \Big(\frac{\widehat \Sigma}{1 + \cos \zeta}\Big)^{\frac{2}{3}} \, \Big[ d\chi^2 ~+~ \frac{(1+\cos \zeta ) \cos^2 \chi }{\widehat \Sigma }\, ds^2_{\IC\IP_2}  \\[6 pt]
& \qquad\qquad  ~+~   \frac{(1+\cos \zeta )^2  }{\widehat \Sigma^2 }\, \sin^2 \chi \,\cos^2 \chi \,\Big(d\psi + \coeff{1}{2}\, \sin^2 \theta\, \sigma_3  +  \frac{2 \, \cos \zeta }{(1+\cos \zeta )} \, d \phi\Big) \Big] \,, 
\end{split}
\end{equation}
The full eleven-dimensional metric limits to the sum of (\ref{3met1}), (\ref{r2met}) and (\ref{6met1}).

Observe that (\ref{6met1}) is conformally K\"ahler.  That is, the metric 
\begin{equation}
\begin{split}
\widehat{ds_6}^2 ~=~ & \frac{\widehat \Sigma }{ (1+\cos \zeta )}\,   d\chi^2 ~+~  \cos^2 \chi \, ds^2_{\IC\IP_2} \\[6 pt]
& ~+~   \frac{(1+\cos \zeta )  }{\widehat \Sigma }\, \sin^2 \chi \,\cos^2 \chi \,\big(d\psi + \coeff{1}{2}\, \sin^2 \theta\, \sigma_3  +  \frac{2 \, \cos \zeta }{(1+\cos \zeta )} \, d \phi\big) \,, \label{6met2}
\end{split}
\end{equation}
has a K\"ahler form:
\begin{equation}\label{Kform1}
\begin{split}
\widehat J ~\equiv~  &- \sin \chi \,\cos \chi \, d\chi \wedge  \Big(d\psi + \frac{1}{2}\, \sin^2 \theta\, \sigma_3  +  \frac{2 \, \cos \zeta }{(1+\cos \zeta )} \, d \phi\Big)  ~+~  \cos^2 \chi \, J_{\IC\IP_2} \\[6 pt]
~=~  & d\, \Big[\coeff{1}{2}\, \cos^2 \chi \, \Big(d\psi + \coeff{1}{2}\, \sin^2 \theta\, \sigma_3  +  \frac{2 \, \cos \zeta }{(1+\cos \zeta )} \, d \phi\Big) \Big]
 \,, 
\end{split}
\end{equation}
where $J_{\IC\IP_2}$ is the K\"ahler form on $\IC\IP_2$.  Here we are, of course, taking $\zeta$ to be constant at its asymptotic value. One can also easily verify that as $\chi \to \pi/2$ this manifold is smooth, and is locally like the origin of $\IR^6$. 

The only singular parts of the metric occur at $\rho =0$ and at $\chi =0$, where there are orbifold singularities in two different $\IR^2$ planes in (\ref{r2met}) and  (\ref{6met1}) respectively.  As we will discuss below, these loci represent the intersections of the various branes that are present in the infra-red limit.

The non-zero components of the Maxwell field are given by:
\begin{equation}
A_{(3)}  ~\sim~  h_0(r,\chi) \, e^1\wedge e^2 \wedge e^3 ~+~  \frac{1}{4} \, \sin \zeta  \, (e^6\wedge e^9 +e^7\wedge e^8  - e^5\wedge e^{10}  )\, \wedge \hat e^{11} \,,\label{A3asymp1}
\end{equation}
with
\begin{align}
 h_0 ~\sim~  & {\rm sign}( 1- 2\cos\zeta) \,\frac{(\cos \zeta - \cos 2\chi)}{\widehat \Sigma}   \, 
\,, \label{h0asymp1} \\[6 pt]
 \hat e^{11} ~\equiv~&     \frac{2\, m_7}{(1 + \cos \zeta)^{\frac{1}{3}}  \,  \widehat \Sigma^{\frac{2}{3}}  }\,  \big(d \phi + \cos^2 \chi \,(d\psi + \coeff{1}{2}\, \sin^2 \theta\, \sigma_3) \big)  \,. \label{e11hatdefn}
\end{align}
Thus $A_{(3)}$ has regular coordinate components.  One might be concerned that the Maxwell tensor has a singular source at $\rho =0$ because the $e^{11}$ is vanishing.  However, the frame components of the Maxwell tensor are, in fact, regular.  The non-zero frame components in the compactified directions (including $e^{11}$) are:
\begin{align}
F_{46911} \,,  F_{47811}  ~\sim~  &- \frac{2\, m_7\,{\rm sign}( 1- 2\cos\zeta) }{  \widehat \Sigma^\frac{4}{3}} \, (1+ \cos  \zeta)^\frac{1}{3}\,(\cos2 \chi  \cos  \zeta -\sin ^2\chi )\,\tan \coeff{1}{2} \zeta   \,, \label{Fasymp1a}\\[6 pt]
F_{451011}  ~\sim~   & -\frac{2\,m_7\, {\rm sign}( 1- 2\cos\zeta)\, \sin \zeta }{ (1+ \cos  \zeta)^\frac{2}{3} \, \, \widehat \Sigma^\frac{1}{3}} \,, \label{Fasymp1b}\\[6 pt]
F_{56910}  \,,   F_{57810}  ~\sim~ & - \frac{2\,m_7\, (1+ \cos  \zeta \,\sin ^2\chi )\, \sin \zeta }{ (1+ \cos  \zeta)^\frac{2}{3} \, \, \widehat \Sigma^\frac{4}{3}} \,, \qquad F_{6789}  ~\sim~    \frac{2\,m_7 \sin \zeta }{(1+ \cos  \zeta)^\frac{2}{3} \, \, \widehat \Sigma^\frac{1}{3}} \,.\label{Fasymp1c}
\end{align}

It is also useful to note that the electric part of the Maxwell field is extremely simple
\begin{equation}
F_{(4)}^{\rm electric} ~=~   d A^{\rm e}_{(3)}  \,, \qquad  A^{\rm e}_{(3)}  ~=~ -\coeff{1}{2} \, R^6\,{\rm sign}( 1- 2\cos\zeta)\,   \cos \coeff{1}{2} \zeta  \, \cos 2 \chi \, dx_1 \wedge dx_2 \wedge dx_3 \label{IRFsimp1}\,.
\end{equation}
%

\subsection{$\zeta = \pm \pi/3 $}
\label{zetaspec}

Here we simply take $\zeta = + \pi/3 $ because the flow for $\zeta = - \pi/3 $ simply involves reversing the sign of the internal components of the flux, $A_{(3)}$.

One now has rather different asymptotics:
\begin{equation}
d \lambda  ~\sim~ \mp \frac{g}{2 \,\sqrt{2}} \, e^{ \lambda}\, dr  \,, 
 \qquad e^A  ~\sim~ R^2 \,  e^{-3\, \lambda}   \label{11dasymp5x}\,,
\end{equation}
\begin{equation}\label{11met1}
\begin{split}
ds_{11}^2  ~\sim~ 2^{-\frac{4}{3}}\, 3^\frac{1}{3}  \,\widehat \Sigma^\frac{2}{3}\, \Big[\,\frac{d\rho^2}{\rho^2} & ~+~  \rho^2 \, R^2\,(-dx_1^2 +dx_3^2 + dx_3^2)   \\[6pt]  
& ~+~ \frac{64}{27  } \,\rho^2 \,  \Big(d \phi + \frac{3\,\cos^2 \chi }{2\,\widehat \Sigma}\,(d\psi + \coeff{1}{2}\, \sin^2 \theta\, \sigma_3 +  \coeff{2}{3} \, d \phi) \Big)^2   \\[6pt]
& ~+~ \frac{4}{3} \,m_7^{-2}\, \Big( \, d \chi^2 ~+~ \frac{3 }{2\,\widehat \Sigma }\,\cos^2 \chi \, ds^2_{\IC\IP_2} \\[6pt]
& \qquad\qquad ~+~   \frac{9 }{4\, \widehat \Sigma^2 }\, \sin^2 \chi \,\cos^2 \chi \,\big(d\psi + \coeff{1}{2}\, \sin^2 \theta\, \sigma_3  +  \coeff{2}{3} \, d \phi\big)\,\Big)^2 \,\Big] \,, \end{split}
\end{equation}
where, as before,
\begin{equation}
\rho ~\equiv~  e^{-2 \lambda} \,, \qquad   \widehat \Sigma \ ~\equiv~ (1 - \coeff{1}{2} \, \cos 2\chi)
 \label{redefs1}\,.
\end{equation}
Note that the compact six-dimensional metric in (\ref{11met1}) is simply the metric (\ref{6met1}) specialized to $\zeta = \pi/3 $ and is therefore also conformally K\"ahler.

Remarkably, for $\zeta = \pi/3 $ many of the components of $F_{(4)}$ vanish in the infra-red and we find that this limiting Maxwell field is simply given by
\begin{equation}
F _{(4)}~=~   d A_{(3)}^0  \,, \qquad  A_{(3)}^0  ~=~ \frac{\sqrt{3} \,m_7^{-3}}{4\, \widehat \Sigma} \,  \cos^4 \chi \ J_{\IC\IP_2}\wedge (d\psi + \coeff{1}{2}\, \sin^2 \theta\, \sigma_3 +  \coeff{2}{3} \, d \phi)
 \label{IRFsimp2}\,.
\end{equation}
Note that the space-time components parametrized by $h_0$ vanish in this limit and that $F$ is purely magnetic and lives entirely on the conformally K\"ahler six-manifold.  Thus, for $\zeta = \pi/3 $ there are only M5 branes in the infra-red:  the M2 branes have dissolved completely.

\subsection{The IR limit of the flows}
\label{IRlimit}

The first and rather remarkable surprise is that the warp factor, $ \DX^{1\over 6}\Sigma^{1\over 3}\,e^{A}$,  in front of frames parallel to the M2-branes  (\ref{d11frames})  is not singular in the infra-red for $\zeta \ne 0, \pi$.  For  $\zeta = 0, \pi$,   this warp factor is expected to be singular because such a flow has no internal fluxes and the warp factor is then simply a power of the harmonic function describing M2 brane sources that have spread on the Coulomb branch.  However (\ref{11dasymp3}) shows that there is no singularity for generic $\zeta$ and for $\zeta =\pm \pi/3 $ equation (\ref{11met1}) shows that this warp factor actually vanishes.  Thus there are no strongly singular sources of M2 branes in the infra-red.  

The second surprise is that the internal six-dimensional manifold goes to a {\it finite-sized} conformally  K\"ahler, six-dimensional manifold and this manifold is smooth at  $\chi = \pi/2$.    Indeed, the only singularities are conical and occur at $\rho =0$ where the $\rm U(1)$ fiber defined by $e^{11}$ pinches off (see (\ref{r2met})) and at $\chi =0$ where  the $\rm U(1)$ fiber defined by $e^{10}$ pinches off (see (\ref{6met1})).  It is also evident from (\ref{h0asymp1}) and (\ref{Fasymp1a})--(\ref{Fasymp1c}) that the core of this holographic flow is populated by finite, smooth electric (M2-brane) and magnetic (M5-brane) fluxes.  Thus there is evidently brane polarization and a geometric transition in which the M2 branes partially dissolve into smooth M5-brane fluxes leaving a finite sized ``bubble'' in the form of a six-dimensional K\"ahler manifold.  This is rather reminiscent of the kind of transition one finds in microstate geometries \cite{Bena:2005va,Berglund:2005vb,Bena:2007kg}.

To understand the brane content in the infra-red in more detail it is perhaps easiest to examine the projectors that define the supersymmetries.   These are given by \eqref{proj1}, \eqref{proj2inf} and \eqref{proj0inf}.  Define the rotated frames
\begin{align}
\Gamma^{\widehat{4}}&~\equiv~\cos\alpha\, \Gamma^{4}+\sin\alpha\, \Gamma^{5}\,, \qquad
\Gamma^{\widehat{5}}~\equiv~\cos\alpha\, \Gamma^{5}-\sin\alpha\, \Gamma^{4}\,, \\[6 pt]
\Gamma^{\widehat{10}}&~\equiv~\cos\omega\, \Gamma^{10}+\sin\omega\, \Gamma^{11}\,, \qquad
\Gamma^{\widehat{11}} ~\equiv~\cos\omega\, \Gamma^{11}-\sin\omega\, \Gamma^{10}\,, 
\end{align}
where $\alpha =\alpha(r,\chi)$ and $\omega =\omega(r,\chi)$ are functions that depend upon the flow.  The details of these angles and how they flow are given in section \ref{subSec:solsuper} and may also be found in  \cite{Pilch:2015vha}.  Given the other projectors and the fact that  $\Gamma^{1\ldots 11}=1$, one can write  \eqref{proj0inf} as
\begin{equation}
\Pi_0 ~\equiv~  \frac{1}{2}(1+\cos\beta\,\Gamma^{123}+\sin\beta\, \Gamma^{\rm Int}) 
\label{Pi0RG}\,,
\end{equation}
where  $\Gamma^{\rm Int}$ is any one of the following
\begin{equation}
\Gamma^{123 6 9\,\widehat{11}} \,, \qquad \Gamma^{123  7 8\,\widehat{11}} \,, \qquad  \Gamma^{123 \widehat{5} \widehat{10}\,\widehat{11}}  \,, \label{Pi0partsRG}
\end{equation}
This means that the flow represents M2 branes polarizing into three sets of M5 branes that have  $(3+1)$ common directions, those of the M2 branes and one compactified direction, defined by $\hat{e}^{11}$.  This means that the directions transverse to the M5 branes are defined by $\hat e^4$, $\hat e^{10}$ and four of the compact internal directions.  Thus the brane wrapping is crucially determined by $\hat{e}^{11}$ and hence by $\omega$.  

For $\cos 3\zeta \ne -1$ and $\lambda \to \infty$, one has: 
\begin{equation}
\cos\beta  ~=~ \frac{\cos \zeta - \cos \chi }{(1-\cos \zeta \, \cos \chi)}  \,, \qquad \alpha~=~ \omega~=~ \frac{\pi}{2}\,.
\label{betaIR1}
\end{equation}
Thus $\hat{e}^{4}= {e}^{5}$, $\hat{e}^{11}= -{e}^{10}$ and so  $\chi$ lies transverse to all the branes.  Indeed, (\ref{betaIR1}) shows that  $\chi=0$  involves only anti-M2 brane sources and so the conical singularity at this point is not altogether surprising.    The locus $\rho \equiv e^{-2 \lambda} =0$ also defines the location of the residual M2 branes and of some of the M5 branes and thus another conical singularity is not surprising.  All the  M5 branes have a common direction along  $e^{10}$, which is the Hopf fiber in the K\"ahler metric (\ref{6met2}). 

One rather interesting flow involves having $\zeta \to \pi/2$ at infinity.  This does not mean that $\zeta= {\pi}/{2}$ all along the flow; indeed (\ref{Iom2}) takes the value $-\frac{1}{2}$  on such a flow and this implies that as $\lambda \to 0$ one must have  $\zeta \to \arccos (\pm \frac{1}{\sqrt{5}})$.  What makes this flow interesting is that $\rm SU(4)$ symmetry is restored in the infra-red.  In particular, the metric (\ref{6met2}) becomes precisely that of $\IC \IP^3$

As described in \cite{Pilch:2015vha}, the situation is very different for  $\cos 3\zeta \ne -1$. For $\zeta = \pi/3$ and $\lambda \to \infty$, one has\footnote{The solution for $\zeta =- \pi/3 $ simply flips the signs of the internal fluxes and is completely equivalent.}: 
\begin{equation}
\cos\beta  ~=~0 \,, \qquad \omega~=~ 0\,,
\label{h0van}
\end{equation}
and 
\begin{equation}
\cos \alpha ~=~  \frac{  2\cos 2\chi -1 } {2-\cos 2\chi} \,, \qquad \sin \alpha ~=~ -  \frac{ \sqrt{3} \sin 2\chi } {2-\cos 2\chi}\,.
\end{equation}
We now have $\hat{e}^{11}= {e}^{11}$ and so the M5 branes wrap ${e}^{11}$ while ${e}^{10}$ remains transverse to the branes.  More significantly, 
the M2-brane flux now vanishes entirely and all that remains is a very simple non-singular magnetic  (M5-brane) flux (\ref{IRFsimp2}).  The limiting metric (\ref{11met1}) is almost like that of $AdS_5 \times  \cB_6$ where $\cB_6$ is the conformally K\"ahler metric.  The five-dimensional manifold that we label as $\widehat{AdS_5}$ is $AdS_5$ in  Poincar\'e form with one spatial direction compactified and fibered over $\cB_6$. Holographically it suggests that the IR phase is almost a CFT except that one spatial direction has been ``put in a periodic'' box of some fixed scale and that some interactions have been turned on so that this direction becomes non-trivially fibered.  Thus the IR phase is almost a CFT fixed point.

Finally, we would like to note that all the fluxes and most, if not all, of the metric in the IR limit are purely functions of $\chi$.  This however, does not mean that these limits represent solutions to the equations of motion because the $r$ (or $\lambda$) dependence is critical to giving finite terms that survive in the IR limit of the  equations of motion.

\section{Generalizations}
\label{Sect:General}

There are several natural generalizations of the results presented here.  The first and most obvious is to use a somewhat more general gauged supergravity Ansatz.  Our Ansatz may be  thought of as reducing to $\Neql2$ supergravity coupled to one vector and with a holomorphic superpotential, $\mathcal{V} ~=~ \sqrt{2} (1+ z^3)$  
\cite{Bobev:2013yra, Pilch:2015vha}.  This can easily be generalized to $\Neql2$ supergravity coupled to three vector multiplets while still remaining within gauged $\Neql8$ supergravity.  This truncation was considered in \cite{Freedman:2013oja} and the holomorphic superpotential becomes
\begin{equation}
\mathcal{V} ~=~ \sqrt{2}\, (1+ z_1 z_2 z_3) \,,
\label{holV2}
\end{equation}
where the $z_i$ are the complex scalars of the three vector multiplets.  Our results here may be thought of as the special case with the three vector multiplets set equal and, in particular, $z_1=z_2=z_3=z$.  As noted in \cite{Pilch:2015vha},  the uplift formulae will be far more complicated, but one expects that the infra-red limit will involve a more general K\"ahler manifold with a $U(1)^3$ symmetry.  It may also have some non-trivial moduli in that the $\zeta = \pi/3$ condition may simply become a constraint on the overall phase of $z_1 z_2 z_3$.  These  moduli would probably be related to the three distinct sets of M5-brane fluxes on the K\"ahler manifold. We intend to investigate this further. 

We have also attempted a much greater generalization in the spirit of  \cite{Pilch:2004yg,Nemeschansky:2004yh,Gowdigere:2005wq}.  One starts with the uplifted flow solution and introduces rotated frames that subsume the need for the rotation by $\alpha$:   
\begin{equation}
\hat e^{4} ~\equiv~\cos\alpha\, e^{4}+\sin\alpha\, e^{5}\,, \qquad \hat e^{5} ~\equiv~-\sin\alpha\, e^{4}+\cos\alpha\, e^{5}\,.
\end{equation}
Rather remarkably one can integrate these frames in our flow solution to find new variables, $(u,v)$,  so that
\begin{equation}
\label{uvframes}
  \hat e^{4} ~\sim~ du \,, \qquad \hat e^{5} ~\sim~ dv\,.
\end{equation}
One can also find explicit expressions for these new coordinates: 
\begin{equation}
u ~\equiv~ e^{2 A} \,p(r) \,  (2 \cos 2 \chi -1) \,, \qquad v ~\equiv~ e^{2 A} \cos^3 \chi  \sin \chi\,.
\end{equation}

Combined with the observation implicit in the simple  and very canonical form of $\hat e^{10}$ given in (\ref{e10hatsimp}), one finds that our flow solution has a some extra structure and, in particular, the pairing of $v$ and the frame in $\hat e^{10}$ in the supersymmetry projectors, along with the phase dependence of the supersymmetries is very suggestive of an underlying six-dimensional complex structure.  

We therefore start with the $(u,v)$ coordinates and their associated frames.  We then take the metric and fluxes to have a completely general \suthuu-invariant Ansatz involving arbitrary functions  of $(u,v)$ everywhere possible.  We  also assume that the frame $\hat e^{10}$ is universal and, in particular, take the metric Ansatz to be of the form:
\begin{equation}\label{gend11frames}
\begin{split}
e^i & \eql H_0(u,v)^{1\over 3}\, f^i\,,\quad i\eql 1,\ldots,3\,,\qquad 
e^4  \eql H_0(u,v)^{-{1\over6 }}\, H_1(u,v) \, du\,, \\[6 pt]
e^5 & \eql H_0(u,v)^{-{1\over6 }}\, H_2(u,v) \,  dv \,,\qquad
e^6  \eql H_0(u,v)^{-{1\over6 }}\, H_3(u,v) \, v \, d\theta \,,\\[6 pt]
e^7 & \eql   \frac{1}{2}\, H_0(u,v)^{-{1\over6 }}\, H_3(u,v) \, v \,   \sin \theta \,  \sigma_1 \,;     \qquad
e^8   \eql  \frac{1}{2}\, H_0(u,v)^{-{1\over6 }}\, H_3(u,v) \, v \,  \sin \theta \, \sigma_2 \,,\\
e^9 & \eql   \frac{1}{2}\, H_0(u,v)^{-{1\over6 }}\, H_3(u,v) \, v \,   \sin \theta \,\cos \theta \, \sigma_3 \,;   \\[6 pt]
e^{10} & \eql   H_0(u,v)^{-{1\over6 }}\, H_4(u,v)  \,\big(d\phi ~+~ \coeff{3}{2}\,(d\psi + \coeff{1}{2}\, \sin^2 \theta\, \sigma_3)  \big) \,;  \\[6 pt]
e^{11} & \eql H_0(u,v)^{-{1\over6 }}\, H_5(u,v) \,  \Big(d \phi + G(u,v) \,\big(d\phi ~+~ \coeff{3}{2}\,(d\psi + \coeff{1}{2}\, \sin^2 \theta\, \sigma_3)  \big)\Big) \,,
\end{split}
\end{equation}
where the $H_a$ and $G$ are, {\it ab initio}, undetermined functions.  For the supersymmetry projectors we  take  $\alpha = \omega =0$ at the outset but retain  $\beta = \beta(u,v)$.  We also assume that the supersymmetries have the same  $\psi$- and $\phi$-dependence as in (\ref{Mphpseqs}).  Note that this implies that the supersymmetry only depends on the combination $\phi+ \frac{3}{2} \psi$, which appears in $e^{10}$.  

For the Maxwell field, we take the most general \suthuu-invariant  potential and choose a gauge in which all the components along $e^4$ vanish:
\begin{equation}
\begin{split}
A^{(3)} ~=~ & h_0 \, e^1 \wedge e^2 \wedge e^3  ~+~ p_0 \, e^5 \wedge e^{10} \wedge e^{11}~+~ p_1 \, e^5 \wedge (e^{6} \wedge e^{9} + e^{7} \wedge e^{8} ) \\[6 pt] 
& ~+~ p_2 \,  (e^{6} \wedge e^{9} + e^{7} \wedge e^{8} ) \wedge e^{10} ~+~ p_3 \,  (e^{6} \wedge e^{9} + e^{7} \wedge e^{8} ) \wedge e^{11} \,,
\end{split}
\end{equation}
where the $h_0$ and the $p$'s are also functions of $(u,v)$.  The whole point is that this Ansatz at least contains our uplifted flow solution and we want to use this more general structure to understand the underlying geometry and perhaps find more general supersymmetric solutions.

To that end, one solves the  supersymmetry variations to fix as many functions as possible.  In   \cite{Pilch:2004yg,Nemeschansky:2004yh,Gowdigere:2005wq} the whole problem reduced to determining a  single ``master function'' from which every other flux and metric function was derived.  This master function itself satisfied a non-trivial, non-linear differential equation.  In spite of the nice structure that we have discovered, the same kind of procedure applied here does not lead to such a simple reduction. We did, however, discover some rather general results that we will briefly summarize so as to give a flavor of what emerges.   

First, we find that, {\it along the flow,}  the metric functions $H_1, H_2$ and $H_4$ are all fixed algebraically in terms of $H_3$ and $H_5$.  Moreover, the $v$ derivative of $H_3$ is simply related to $H_3$ and $H_5$.  These conditions combine to reveal that the non-compact, eight-dimensional metric transverse to the M2-branes, $ds_8^2$,  has six-dimensional foliations, defined by the $\IC \IP^2$, the coordinate $v$ and the frame $ e^{10}$,  are necessarily a $u$-dependent family of K\"ahler metrics.  By this we mean that there is indeed a K\"ahler form, $J$, on each leaf of the foliation but $d J$ is  proportional to $du$:  If $u$ were held constant, $d J$ would indeed vanish. Thus the six-dimensional  K\"ahler structure apparent in the IR actually descends from a family of such structures along the flow.

Secondly, we find that the flux parametrized by $p_0$ is necessarily pure gauge and so we can set $p_0 \equiv 0$ without loss of generality.    The rest of the $p$'s satisfy a complicated system of equations that link them with $\beta$ and the remaining metric functions.  Ultimately one can reduce the system to show that all the unknown functions are determined in terms of {\it two} unknown functions of $(u,v)$, $H_3$ and  a function that we will call $F$.    The former satisfies an extremely complicated non-linear differential equation while $F$ is a pre-potential in that its derivatives determine some of the functions in the Ansatz:    The the $v$-derivative of $F$ gives the function $G$ in (\ref{gend11frames}) while the $u$-derivative of $F$ gives $H_5^{-2} \cos \beta$.  The remarkable fact is that $F$ is     {\it harmonic} (annihilated by the Laplacian) in a metric that is conformal to $ds_8^2$.   The  Laplacian on $ds_8^2$ explicitly involves $H_3$  and so the hamonicity of $F$  is far from simple to use in practice.  

We have, of course, verified that the uplifted flows do indeed satisfy these conditions but so far have not been able to simplify and elucidate the general discussion to the degree that it is worthy of presentation in this paper.   The important bottom-line though is that every generalization we have considered shows  the same structure for the eight-manifold transverse to the M2 branes: it consists of six-dimensional K\"ahler manifolds foliated over an $\IR^2$ base where one of the $\rm U(1)$'s acts as an isometry and the  K\"ahler potentials depend upon the radial coordinate in this $\IR^2$.

\section{Conclusions}
\label{Sect:conclusions}

On a technical level, our results represent a highly non-trivial test of  the recent results on uplifting gauged $\Neql8$ supergravity to M theory.  Indeed, our discussion in Section \ref{Sect:General} illustrates just how difficult it would be to construct the very symmetric class of flows we consider directly within eleven dimensions.  

More broadly, we have, once again, seen how apparently very singular ``Flows to Hades'' in gauged supergravity can encode some very interesting physical flows and Janus solutions when lifted to M theory.  Results like this illustrate why it is very important to understand how gauged supergravities are encoded in higher dimensional theories.

The uplift formulae to eleven-dimensional supergravity have been well-studied and tested compared to the uplift of gauged $\Neql 8$ supergravity in five dimensions to IIB supergravity in ten dimensions.  One reason might be that for quite some time   only  uplift formulae for the internal metric and dilaton were known \cite{Pilch:2000ue}  and those were inferred by analogy to the M-theory result rather than proven directly.  However, this has changed recently with various reformulations of type IIB supergravity and the resulting uplift formulae for all fields \cite{Lee:2014mla,Ciceri:2014wya,Baguet:2015sma}. 
 While there are quite a number of interesting physical examples of IIB flows, it is possible that there are others yet to be discovered because they look singular from the five-dimensional perspective. Given that the dual theory in $\Neql 4$ Yang-Mills theory, it would be interesting to examine such flows using the new Ans\"atze  that are now available.

The limitation in using gauged supergravity is that one is restricted to a relatively small family of fields from the higher-dimensional perspective.  Fortunately the fields one has are quite a number of the simplest relevant and marginal perturbations and so one can still probe interesting physics.  The limitation is most sorely felt when one tries to probe details and subtleties of the families of IR fixed points and for this gauged supergravity is too blunt an instrument.  On the other hand, the uplifts of gauged supergravity solutions can give invaluable insights into the  geometric structures that underlie the more general classes of flow and thus enable broader, and perhaps more physically interesting solutions to be found.  This was  evident for  \nBPS{2} flows \cite{Pope:2003jp,Bena:2004jw,Lin:2004nb}  and so even though gauged supergravity sometimes does not describe the exact physics one wants, it can motivate and inform the search for  physically interesting families of solutions.

 In this spirit, we suspect that the families of flows and Janus solutions considered here should admit interesting generalizations.   There are the generalizations within gauged supergravity as outlined at the beginning of Section \ref{Sect:General}.  However, there should be families that involve a six-dimensional K\"ahler manifold fibered over a two-dimensional base with a $\rm U(1)$ isometry.  It will probably be very challenging to use merely this information to find general flow solutions.  However, we saw here that the fluxes also took on a relatively simple form and if one could understand the geometry underlying this one should be able to move towards the general class of solutions.  
 
From the physical perspective, the flows and Janus solutions we have constructed are very interesting in that they involve M2 branes polarizing and dissolving into {\it non-singular} (except for orbifolds) distributions of M5 branes and for one choice of parameter, flowing to a higher-dimensional, almost-conformal fixed point \cite{Pilch:2015vha}.  Indeed, it was shown in \cite{Pilch:2015vha} that such flows exist for any choice of parameter if one does not insist on supersymmetry.   

Apart from the interesting holographic interpretation of these flows, this kind of mechanism also underpins the microstate geometry program   in which black holes are replaced with smooth, horizonless solitonic geometries. (For reviews, see \cite{Bena:2007kg,Bena:2013dka,Gibbons:2013tqa}.)  The basic mechanism means that the black hole undergoes a phase transition (driven by the Chern-Simons interactions) in which the singular electric charge sources are replaced my smooth magnetic fluxes. In M~theory, this is realized by M2 charges being replaced by M5 fluxes.  There are vast families of supersymmetric examples of this in asymptotically flat backgrounds but so far there are no examples of this process (supersymmetric, or not)  in asymptotically  $AdS_4$ or $AdS_5$ space-times. It is thus extremely helpful to have an example of such a transition and perhaps use it to understand how such a phase transition might occur more generally in asymptotically  $AdS_4$ or $AdS_5$ space-times.
 
Finally, there is the question of whether there are IIB analogs of the flows  and Janus solutions studied here.   These would probably be flows in which D3 branes polarized into families of intersecting D5 branes while preserving $\Neql 1$ supersymmetry in the field theory along the flow. Such solutions  might even ``flow up dimensions'' to give a compactified higher-dimensional field theories in the infra-red.

\section*{Acknowledgments}

We would like to thank N.~Bobev,  H.~Godazgar, M.~Godazgar, O.~Kr\"uger and H.~Nicolai for helpful discussions and correspondence. This work was supported in part by  DOE~grant DE-SC0011687.

\appendix
\section{Conventions} 
\label{Appendix:A}

We use the same conventions as in \cite{Gowdigere:2003jf} with the ``mostly plus metric'' and the eleven-dimensional  equations of motion given by
\begin{equation}\label{Eineqs11}
R_{MN}+g_{MN}R  \eql {1\over 3}\,F_{MPQR}F_{N}{}^{PQR}\,,
\end{equation}
\begin{equation}\label{Maxeqsd11}
\nabla_MF^{MNPQ}\eql -{1\over 576}{1\over\sqrt{-g}}\,\epsilon^{NPQR_1\ldots R_8}F_{R_1\ldots R_4}F_{R_5\ldots R_8}\,.
\end{equation}
 The Maxwell equation can be rewritten in terms of forms as\footnote{Note that our normalization of the four-form flux is according to the ``old supergravity convention.''}
\begin{equation}\label{maxwell}
d\star F_{(4)}+F_{(4)}\wedge F_{(4)}\eql 0\,,
\end{equation}
where $\star\equiv *_{1,10}$ is the Hodge dual in eleven dimensions. 

In general, we define the Hodge dual of a $k$-form, $\omega$, in $d$-dimensions by
\begin{equation}\label{}
(*\,\omega)_{i_1\ldots i_{d-k}}\eql {1\over k!}\, \eta_{i_1\ldots i_{d-k}}{}^{j_1\ldots j_k}\,\omega_{j_1\ldots j_k}\,,
\end{equation}
where 
\begin{equation}\label{}
\eta^{i_1\ldots i_d}~\equiv~ {1\over \sqrt{|g|}} \,\epsilon^{i_1\ldots i_d}\,,\qquad \epsilon^{i_1\ldots i_d}\eql 1\,.
\end{equation}
Then
\begin{equation}\label{}
(*\,\omega)\wedge\omega\eql \pm |\omega|^2\,{\rm vol}\,,\qquad  |\omega|^2~\equiv~ {1\over k!}\omega_{i_1\ldots i_k}\omega^{i_1\ldots i_k}\,.
\end{equation}
with the $+$ sign is for a positive definite metric and the $-$ sign for a Minkowski signature mostly plus metric.
 
For a $(p,q)$-form $\Omega_{(p,q)}$ on $\cals M_{1,3}\times \cals M_7$ with the warped product metric \eqref{d11met},   we have a convenient  decomposition of the Hodge dual:
\begin{equation}\label{decdual}
\star\Omega_{(p,q)}\eql (-1) ^{p(7-q)} *_{1,3}*_{7}\,\Omega_{(p,q)} \,,
\end{equation}
where $*_{1,3}$ and  $*_7$  are, respectively, the dual on  $\cals M_{1,3}$ with respect to the four-dimensional part of the metric, $g_{\mu\nu}$, and   the dual on $\cals M_7$ with the internal metric, $g_{mn}$. Factoring out the warp factor, we have
\begin{equation}\label{stdual}
*_{1,3} \omega_{(p)}\eql \Delta ^{p-2}\sto_{1,3} \omega_{(p)}\,,
\end{equation}
where $\omega_{(p)}$ is a $p$-form on $\cals M_{1,3}$ and $\sto_{1,3}$ is the dual with respect to $\go_{\mu\nu}$.

\section{Reduced $\rm E_{7(7)}$ tensors and the scalar action} 
\label{Appendix:B}

\def\AA#1#2#3#4{A_{2\,#1}{}^{#2#3#4} }
\def\cA{\mathcal{A}}

Using the 56-bein \eqref{56bein} and with the $\rm SO(8)$ gauge field set to zero,  the $\rm SU(8)$~composite gauge field of the $\Neql8$ theory, 
\begin{equation}\label{atens}
\cA_\mu{}^{ijkl}~\equiv~\cA_\mu{}^{[ijkl]}\eql -2\sqrt 2\,\left(u^{ij}{}_{IJ}\partial_\mu v^{klIJ}-v^{ijIJ}\partial_\mu u^{kl}{}_{IJ}\right)\,,
\end{equation}
has the following non-vanishing components:
\begin{equation}\label{}
A_\mu{}^{1234}\eql A_\mu{}^{1256}\eql A_\mu{}^{1278}\eql A_\mu{}^{3456}\eql A_\mu{}^{3478}\eql A_\mu{}^{5678}\eql -{\sqrt 2\,\partial_\mu \bar z\over 1-|z|^2}\,.
\end{equation}
Hence the kinetic action of the scalar fields is
\begin{equation}\label{}
\begin{split}
e^{-1}\cals L_{kin}  ~\equiv~ -{1\over 96}\, \cals A_{\mu \,ijkl}\cals A_\mu{}^{ijkl} & \eql -3\,{\partial_\mu z\partial^\mu\bar z\over (1-|z|^2)^2}\\[6 pt]
& \eql -3\,\partial_\mu\lambda\partial^\mu\lambda -{3\over 4}\sinh^2(2\lambda)\,\partial_\mu\zeta\partial^\mu\zeta
\,.
\end{split}
\end{equation}

Similarly, for  the $A$-tensors, $A_1^{ij}\equiv A_1^{ji}$ and $A_{2i}{}^{jkl}\equiv A_{2i}{}^{[jkl]}$, we have:
\begin{equation}\label{A1tens}
A_1^{11}\eql\ldots\eql A_1^{66}\eql {1+ z\bar z^2\over (1-|z|^2)^{3/2}}\,,\qquad A_1^{77}\eql A_1^{88}\eql {1+z^3\over  (1-|z|^2)^{3/2}}\,,
\end{equation}
and
\begin{equation}\label{}
\begin{split}
\AA1234 & \eql \AA1256\eql \AA3124\eql \AA3456\eql \AA5126\eql \AA5346\eql -{(1+z)\bar z\over (1-|z|^2)^{3/2}}\,,\\[6 pt]
\AA2134 & \eql \AA2156\eql \AA4123\eql  \AA4356\eql \AA6125\eql \AA6345\eql {(1+z)\bar z\over (1-|z|^2)^{3/2}}\,,\\[6 pt]
\AA 1278 &\eql \AA3478\eql \AA 5678\eql-{(1+z)z\over (1-|z|^2)^{3/2}}\,,\\[6 pt] 
 \AA2178 & \eql \AA4378\eql \AA6578\eql  {(1+z)z\over (1-|z|^2)^{3/2}}\,,\\[6 pt]
\AA7128 & \eql  \AA7348\eql \AA7568\eql -{z+\bar z^2\over (1-|z|^2)^{3/2}}\,,\\[6 pt]
\AA8127 & \eql\AA8347\eql \AA 8567\eql {z+\bar z^2\over (1-|z|^2)^{3/2}}\,.
\end{split}
\end{equation}
Then the scalar potential is
\begin{equation}\label{}
\cals P~\equiv~-\big(\coeff{3}{4}\left|A_1{}^{ij}\right|^2-\coeff{1}{24}\left|A_{2i}{}^{jkl}\right|^2\big)\eql -{6(1+|z|^2)\over 1-|z|^2}\eql -6\cosh( 2\lambda)\,.
\end{equation}

\section{The Freund-Rubin flux} 
\label{Appendix:FR}
 
The calculation of the space-time part of the flux, $F_{(4)}^{\rm st}$, using the method employed  in section~\ref{subSec:stflux} is quite involved even for much simpler solutions such as  uplifts of stationary points. For the latter solutions only the Freund-Rubin part of the space-time flux is present, so that 
\begin{equation}\label{}
F_{(4)}^{\rm st}\eql  \frak{f}_{\rm FR}\, {\rm v\oo l}_{1,3} \,,
\end{equation}
is proportional to the volume of the four-dimensional space-time, $\cals M_{1,3}$, where the proportionality constant  is determined universally by the scalar potential of the four-dimensional theory    \cite{Nicolai:2011cy}. This has been generalized recently in \cite{Godazgar:2015qia} to uplifts of arbitrary solutions by including corrections proportional to derivatives of the scalar potential.
The new conjectured formula for the Freund-Rubin flux, $\frak{f}_{\rm FR}$, reads
\begin{equation}\label{FRflux}
\frak{f}_{\rm FR}\eql {m_7\over 2} \Big[ \,\cals P-{1\over 24}\Big(Q^{ijkl}\widehat\Sigma_{ijkl}+\text{h.c.}\Big)\,\Big]\,.
\end{equation}
The  $Q^{ijkl}$  tensor  is proportional to the first variation  of the potential, $\cals P$, along the noncompact  generators of $\rm E_{7(7)}$ acting on the scalar coset, $\rm E_{7(7)}/SU(8)$. It is given by \cite{deWit:1983gs}
\begin{equation}\label{}
Q^{ijkl}\eql {3\over 4}\,A_{2\,m}{}^{n[ij}A_{2\,n}^{kl]m}-A_1{}^{m[i}A_{2\,m}{}^{jkl]}\,.
\end{equation}
The second tensor in  \eqref{FRflux} is a self-dual contraction
\begin{equation}\label{}
\widehat\Sigma_{ijkl}\eql (u_{ij}{}^{IJ}u_{kl}{}^{KL}-v_{ijIJ}v_{klKL})K^{IJKL}\,,
\end{equation}
where
\begin{equation}\label{}
K^{IJKL}\eql \go^{mn}K_{m}^{[IJ}K_n^{KL]}\,.
\end{equation}

Note that at a stationary point of the scalar potential, the $Q$-tensor  becomes anti-self-dual \cite{deWit:1983gs} and hence the contraction terms in \eqref{FRflux} vanish.

Specializing  the contraction in \eqref{FRflux} to the present solution we find
\begin{equation}\label{}
Q^{ijkl}\widehat \Sigma_{ijkl}+\text{h.c.}\eql -16\,\xi\, \sinh(2\lambda)\cos\zeta\,.
\end{equation}
Then, using  \eqref{d4Pot}, \eqref{xifunct} and \eqref{UandV}, we obtain
\begin{equation}\label{}
\begin{split}
\frak{f}_{\rm FR} & \eql {m_7\over 2}\,\Big[  -6 \cosh(2\lambda)+2(1-4\sin^2\chi)\sinh(2\lambda)\cos\zeta\Big]\\[6 pt]
& \eql {m_7\over 3}\, U\,,
\end{split}
\end{equation}
which agrees with the calculation of the space-time flux in section~\ref{subSec:stflux}.

\section{The Ricci tensor} 
\label{Appendix:C}

The  non-vanishing coefficients of the diagonal components of the Ricci tensor, $R_{MM}$, as defined in \eqref{ricform}: 
\begin{equation}\label{}
\begin{split}
\cals A_{1}& \eql  {1\over 6 \DX^{13/3} \Sigma^{8/3}\Xi^2} \Big[
  \left(p^2+1\right)   \Sigma    X \left(p^2+X^2+1\right)-8 \left(p^2+1\right)^2 X^2\\[6 pt]
 & \hspace{130 pt} +\Sigma ^2
   \left(3 \left(p^3+p\right)^2+3 p^2 X^4-2 \left(3 p^4+7 p^2+4\right) X^2\right)\Big]\\[10 pt]
\cals B_{1} & \eql     {\DX^2\over 6 \DX^{13/3} \Sigma^{8/3}\Xi^2} \Big[
-8 p^2 X^2+16 p^2 \Sigma  X \\[6 pt]
& \hspace{130 pt}+\Sigma ^2 \left(3 p^4-2 p^2 \left(3 X^2+1\right)+3
   \left(X^2-1\right)^2\right)\Big]\,,\\[10 pt]
\cals C_{1} & \eql  {p\DX\over 3 \DX^{13/3} \Sigma^{8/3}\Xi^2}
\Big[-4 \Sigma  X \left(3 p^2+X^2+3\right)+8 \left(p^2+1\right) X^2\\[6 pt] 
&\hspace{130 pt}+\Sigma ^2 \left(-3 p^4+p^2
   \left(6 X^2-2\right)-3 X^4+10 X^2+1\right)\Big]\,,\\[10 pt]
\cals D_{1} & \eql  { 2m_7^{-2}\over   3\DX^{4/3}\Sigma^{8/3}}\,\Big[\Sigma ^2 \left(6 X^2-g^2 m_7^2 \left(3 p^2+3 X^2+4\right)\right)\\[6 pt] &\hspace{130 pt}-2 \Sigma  X \left(g^2
   m_7^2+2 p^2+2\right)-2 \left(p^2+1\right) X^2
\Big]\,;
\end{split}
\end{equation}

\begin{equation}\label{}
\begin{split}
\cals A_{4} & \eql {1\over 3 \DX^{13/3}\Sigma^{8/3}\Xi^2}
\Big[2 \left(p^2+1\right)^2 \left(3 p^2-1\right) X^2-2 \left(3 p^4+2 p^2-1\right) \Sigma  X
   \left(p^2+X^2+1\right)\\[6 pt] & \hspace{180 pt}+\Sigma ^2 \left(3 \left(p^3+p\right)^2+3 p^2 X^4-2
   \left(p^2+1\right) X^2\right)\Big]\,,
   \\[10 pt]
\cals B_{4} & \eql {1\over 3 \DX^{7/3}\Sigma^{8/3}\Xi^2}\Big[-2 p^2 \Sigma  X \left(3 p^2+3 X^2-5\right)+2 \left(3 p^2-1\right) p^2 X^2\\ & \hspace{200 pt}
+\Sigma ^2
   \left(3 p^4-2 p^2+3 \left(X^2-1\right)^2\right)
\Big]\\[10 pt]
\cals C_{4} & \eql {1\over 3 \DX^{10/3}\Sigma^{8/3}\Xi^2}
\Big[2 \Sigma  X \left(3 p^2 \left(p^2+X^2\right)+X^2-3\right)-2 \left(3 p^4+2 p^2-1\right)
   X^2\\[6 pt] & \hspace{200 pt}+\Sigma ^2 \left(-3 p^4-2 p^2-3 X^4+4 X^2+1\right)
\Big]\,,\\[10 pt]
\cals D_{4} & \eql {m_7^{-2}\over 3\DX^{4/3} \Sigma^{8/3}}\,\Big[
2 \Sigma ^2 \left(-3 m_7^2 g^2 \left(p^2+X^2\right)-
4 m_7^2 g^2+6 X^2\right)\\[6 pt] & \hspace{200 pt}-4 \Sigma  X
   \left(m_7^2 g^2+2 p^2+2\right)-4 \left(p^2+1\right) X^2
\Big]\,;
\end{split}
\end{equation}

\begin{equation}\label{}
\begin{split}
\cals A_5 & \eql-{4\over 3\DX^{10/3}\Sigma^{8/3}\Xi^2} \,\left(p^2+1\right) (\Sigma -X) \left(p^2-\Sigma  X+1\right)\,,
\\[10 pt]
\cals B_5 & \eql {4p^2 X (\Sigma -X)^2\over 3\DX^{10/3}\Sigma^{8/3}\Xi^2} \,,
\\[10 pt]
\cals C_5 & \eql -{4p\over 3\DX^{10/3}\Sigma^{8/3}\Xi^2}\,(\Sigma -X) \Big[\Sigma  \left(p^2+X^2+1\right)-2 \left(p^2+1\right) X\Big]\,,
\\[10 pt]
\cals D_5 & \eql \frac{2}{3 \Sigma ^{8/3} X^{4/3}} \,\Big[-2 \Sigma  X \left(g^2-4 m_7^2 \left(p^2+1\right)\right)+2 g^2 \Sigma
   ^2+m_7^2 \left(4 p^2+1\right) X^2\Big]\,;
\end{split}
\end{equation}

\begin{equation}\label{}
\begin{split}
\cals A_6 & \eql {2\over 3\DX^{10/3}\Sigma^{8/3}\Xi^2}\,\left(p^2+1\right) (\Sigma -X) \left(p^2-\Sigma  X+1\right)
\,,\\[10 pt]
\cals B_6 & \eql -\frac{2 p^2 (\Sigma -X)^2}{3 \Sigma ^{8/3} X^{7/3} \Xi^2}
\,,\\[10 pt]
\cals C_6 & \eql {2p\over 3\DX^{10/3}\Sigma^{8/3}\Xi^2}\,(\Sigma -X) \Big[\Sigma  \left(p^2+X^2+1\right)-2 \left(p^2+1\right) X\Big]
\,,\\[10 pt]
\cals D_6 & \eql {2\over 3 \DX^{4/3}\Sigma^{8/3}}\,\Big[
\Sigma ^2 \left(m_7^2 \left(9 p^2+6\right)-g^2\right)\\[6 pt] &
\hspace{100 pt}+\Sigma  X \left(g^2+2 m_7^2
   \left(p^2+1\right)\right)+m_7^2 \left(p^2+1\right) X^2\Big]
\,;
\end{split}
\end{equation}

\begin{equation}\label{}
\begin{split}
\cals A_{10} &\eql  {2\over 3\DX^{10/3}\Sigma^{8/3}\Xi^2}\,\left(3 p^4+4 p^2+1\right) (\Sigma -X) \left(p^2-\Sigma  X+1\right)
\,,\\[10 pt]
\cals B_{10} & \eql  {2\,p^2 X\over 3\DX^{10/3}\Sigma^{8/3}\Xi^2}\,  (\Sigma -X) \left[\left(3 p^2+1\right) X+\Sigma  \left(2-3 X^2\right)\right]\,,\\[10 pt]
\cals C_{10} & \eql -{4p\over  3\DX^{10/3}\Sigma^{8/3}\Xi^2}\,
(\Sigma -X) \left(\Sigma  \left(-\left(3 p^2+2\right) X^2+p^2+1\right)+\left(3 p^4+4
   p^2+1\right) X\right)
\,,\\[10 pt]
\cals D_{10} & \eql -{2 \over 3\DX^{4/3}\Sigma^{8/3}}\,\Big[2 \Sigma  X \left(g^2-4 m_7^2 \left(p^2+1\right)\right)-2 g^2 \Sigma ^2-m_7^2 \left(4
   p^2+1\right) X^2
\Big]
\,;
\end{split}
\end{equation}

\begin{equation}\label{}
\begin{split}
\cals A_{11} & \eql {2\over 3\DX^{13/3}\Sigma^{8/3}\Xi^2}\,\Big[
  \left(p^2+1\right)^2 \left(3 p^2-1\right) X^2-  \left(3 p^4+2 p^2-1\right) \Sigma  X
   \left(p^2+X^2+1\right)\\[10 pt]
   & \hspace{180pt}+{\Sigma ^2\over 2} \left(3 \left(p^3+p\right)^2+3 p^2 X^4-2
   \left(p^2+1\right) X^2\right)
\Big]
\,,\\[10 pt]
\cals B_{11} & \eql {1\over 3\DX^{7/3}\Sigma^{8/3}\Xi^2}\,\Big[
2 \left(3 p^2-1\right) p^2 X^2-2 p^2 \Sigma  X \left(3 p^2+3 X^2-5\right)\\[10 pt]
   & \hspace{180pt}+\Sigma ^2
   \left(3 p^4-2 p^2+3 \left(X^2-1\right)^2\right)
\Big]
\,,\\[10 pt]
\cals C_{11} & \eql {2p\over 3\DX^{10/3}\Sigma^{8/3}\Xi^2}\,\Big[
2 \Sigma  X \left(3 p^2 \left(p^2+X^2\right)+X^2-3\right)-2 \left(3 p^4+2 p^2-1\right)
   X^2\\[10 pt]
   & \hspace{180pt}+\Sigma ^2 \left(-3 p^4-2 p^2-3 X^4+4 X^2+1\right)
\Big]
\,,\\[10 pt]
\cals D_{11} & \eql {2\over 3\DX^{4/3}\Sigma^{8/3}}\,\Big[-\Sigma ^2 \left(3 g^2 p^2-g^2 \left(3 X^2+1\right)+6 m_7^2
   \left(X^2+1\right)\right)\\[10 pt]
   & \hspace{180pt}-4 \Sigma  X \left(m_7^2
   \left(p^2+1\right)-g^2\right)-m_7^2 \left(2 p^2-1\right) X^2
\Big]
\,.
\end{split}
\end{equation}
The non-vanishing coefficients of the off-diagonal components of the Ricci tensor, $R_{MN}$, $M\not=N$, as defined in  \eqref{ricoff}:
\begin{equation}\label{}
\begin{split}
\cals A_{45} & \eql- {2m_7\tan\chi\over \DX^{11/6} \Sigma^{8/3}\Xi}\,\left(p^2+1\right) (\Sigma -X) (2 \Sigma +X)\,,\\[10 pt]
\cals B_{45} & \eql - {2m_7\tan\chi\,p\over \DX^{5/6} \Sigma^{8/3}\Xi}\,(X-\Sigma ) (2 \Sigma +X)\,;
\end{split}
\end{equation}
\begin{equation}\label{}
\cals D_{10\,11}\eql -\frac{2 \left(g^2-2 m_7^2\right) \tan (\chi ) (\Sigma -X)}{\Sigma ^{5/3}X^{1/3}}\,. 
\end{equation}

\vfill\eject



\end{document}